\documentclass[fleqn,useAMS, usenatbib]{mnras} 

\usepackage{newtxtext,newtxmath}
\usepackage[T1]{fontenc}
\usepackage{times}

\usepackage{graphicx}	
\usepackage{amsmath}	
\usepackage{tabularx}

\newcommand{\msun}{{\,\rm M_\odot}}

\def\gsim{ \lower .75ex \hbox{$\sim$} \llap{\raise .27ex \hbox{$>$}} }
\def\lsim{ \lower .75ex \hbox{$\sim$} \llap{\raise .27ex \hbox{$<$}} }
\def \MTWOC{M_{\rm 200c}} 
\def \RTWOC{R_{\rm 200c}}

\title[Galaxy stripping in IllustrisTNG]{Numerical effects on the stripping of dark matter and stars in IllustrisTNG galaxy groups and clusters}

\author[M.~R.~Lovell et al.]{Mark ~R.~Lovell$^{1,2,3}$\thanks{E-mail: m.r.lovell@durham.ac.uk},
Annalisa Pillepich$^{4}$, Christoph Engler$^{4}$, Dylan Nelson$^{5}$, Rahul Ramesh$^{5}$, 
\newauthor
Volker Springel$^{6}$, and Lars Hernquist$^{7}$
\\
$^{1}$ Institute for Computational Cosmology, Durham University, South Road, Durham DH1 3LE, United Kingdom\\
$^{2}$ Department of Physics, Durham University, South Road, Durham DH1 3LE, United Kingdom\\
$^{3}$Centre for Astrophysics and Cosmology, Science Institute,  University of Iceland, Dunhaga 5, 107 Reykjav\'ik, Iceland\\
$^{4}$Max-Planck-Institut f\"ur Astronomie, K\"onigstuhl 17, D-69117 Heidelberg, Germany\\
$^{5}$Universit\"at Heidelberg, Zentrum f\"ur Astronomie, Institut f\"ur theoretische Astrophysik, Albert-Ueberle-Str. 2, D-69120 Heidelberg, Germany \\
$^{6}$ Max Planck Institut f\"ur Astrophysik, Karl-Schwarzschild-Stra\ss e 1, D-85748 Garching bei M\"unchen, Germany\\
$^{7}$ Center for Astrophysics $|$ Harvard \& Smithsonian, 6P Garden St., Cambridge, MA 02138, USA
}

\date{Accepted XXX. Received YYY; in original form ZZZ}

\pubyear{2025}

\begin{document}
\label{firstpage}
\pagerange{\pageref{firstpage}--\pageref{lastpage}}
\maketitle

\begin{abstract}

\noindent The stellar haloes and intra-cluster light around galaxies are crucial test beds for dark matter (DM) physics and galaxy formation models. We consider the role that the numerical resolution plays in the modelling of these systems by studying the stripping of satellites in the IllustrisTNG cosmological simulations. We focus on host haloes of total halo mass $\MTWOC=10^{12-15}\msun$ and satellites of stellar mass $>10^{7}$~$\msun$, and compare stellar halo / satellite properties across 9 IllustrisTNG runs with baryonic particle mass resolution between $8.5\times10^4\msun$ and $7\times10^8$~$\msun$, using a Lagrangian-region technique to identify counterpart satellites across different resolution simulations of the same volume. We publish the corresponding catalogues alongside this paper. We demonstrate that the stripping of DM from satellites that orbit in group- and cluster-mass hosts is largely independent of resolution at least until 90 per cent of their initial mass at infall has been stripped. We do not find evidence for spurious disruption of galaxies due to insufficient resolution for the satellite masses we consider. By contrast, the stripping of stellar mass is strongly resolution-dependent: each factor of 8 improvement in particle stellar mass typically adds 2~Gyr to the stripping time. Improved numerical resolution within the IllustrisTNG model generally results in more compact satellites with larger stellar masses, which in turn generate more centrally concentrated stellar haloes and intra-cluster mass profiles. However, the concomitant increase in stellar mass with increased resolution of both satellites and hosts may still be the cause for the overprediction of the stellar halo mass at large host radii relative to observations seen in some previous studies.

\end{abstract}




\section{Introduction}
\label{sec:int}

 The diffuse stellar components of galaxies, galaxy groups and galaxy clusters provide a strong set of challenges that galaxy formation models in the full cosmological setting need to satisfy simultaneously. For example, a series of physical processes and relations have to be modelled accurately in order to obtain the correct rate of stellar mass stripping, and thus to obtain a good match to the measured surface density of stellar haloes \citep{Merritt16,Merritt20,Gilhuly22,Alonso23,Brough24}. There are at least four requirements to attain this goal: (i) accurately predict the abundances and density profiles of the accreted galaxies; (ii) match the mass--size relation of the host galaxy; (iii) predict the stellar mass--dark matter (DM) halo mass relation of both satellites and hosts; and (iv) also capture the degree of adiabatic contraction  \citep{Blumenthal86,Gnedin04} or any other possible baryonic modifications to the properties of the host halo. All of these aspects have to be addressed within the confines of simulations with finite mass resolution and spatial resolution.

Modern cosmological hydrodynamical simulations of galaxy formation \citep{Vogelsberger20} model the constituent components of matter, including DM, stellar populations, gas, and supermassive black holes, as discrete packets of material, known as resolution elements, with a mass of the order of tens of thousands to billions of solar masses. This discretization of what are, at least in practical terms for gas and DM, continuous fluids leads to a series of limitations to the simulations, which we outline below.

First, massive particles have to be simulated with a softened gravitational force law in order to suppress spurious two-body interactions \mbox{\citep{Power03,Springel08b}}. This softened force law artificially lowers the halo density at a length scale of one softening length, and thus a halo can be disrupted more readily and have a different internal structure than it would at `perfect' resolution \citep[][]{Ludlow19,Ludlow20,Ludlow23}. 
Second, changing the number of particles while keeping the subgrid galaxy formation model parameters the same can lead to changes in galaxy properties. \mbox{\citet{Pillepich17}} and \citet{Engler21} showed that the stellar mass formed at fixed halo mass increased by up to a factor of 2 when changing the mass resolution by a factor of 170. \citet{Lovell20} showed on a galaxy-by-galaxy basis that the masses of Eagle-model Local Group dwarfs in different classes of stellar mass could change significantly with a factor of 12 decrease in DM particle mass: increased resolution {\it suppressed} stellar mass for $L_{*}$ galaxies but {\it enhanced} stellar mass production in dwarf galaxies. Moreover, differences between galaxy formation models will lead to different resolution dependencies, making it necessary to assess the resolution dependence of quantities of interest for each model individually.

Further issues are associated with secondary dynamics in the simulations. One such possibility is that typical choices for the relationship between the resolution element mass (i.e. mass resolution) and gravitational softening length (spatial resolution) results in the spurious disruption of subhaloes at all resolutions. Making such a choice is inevitable as a compromise between two-body scattering versus non-Newtonian forces \citep{vanKampen00}. Specifically, \citet{vdBosch18} performed idealised simulations of subhaloes in spherical orbits. They argued that, in the scenario of a subhalo of mass 1000 times smaller than the host mass and in a circular orbit of 10~per~cent of the host radius, the mass of the subhalo ought to tend asymptotically to one thousandth of the initial mass over the age of the Universe at near perfect resolution, but instead is completely disrupted after $\sim5$~Gyr when using common mass resolution--spatial resolution choices. Subsequent work by \citet{Green21} found that the influence of such effects is smaller in the full cosmological model, although their analysis was restricted to a single $N$-body simulation, the Bolshoi simulation \citep{Klypin11}, which resolves the dwarf haloes responsible for stellar halo construction with only 50 particles. In fact, according to the results of \citet{Bahe19} based on Hydrangea, the disruption of satellite galaxies is not a ubiquitous feature of cosmological galaxy cluster simulations, at least not for peak total masses above $10^{10}\msun$ and with $\sim10^6\msun$ ($\sim10^7$) baryonic (DM) mass resolution. However, \citet{Errani23} have argued that at infinite resolution a stripped remnant will persist indefinitely. At the other end of the mass scale, \citet{Grand21} have shown using a very high resolution hydrodynamical simulation of a single $L_{*}$ galaxy that higher resolution makes satellites resilient to stripping on radial orbits, where a factor of 64 improvement in resolution enables satellites to survive orbital impact parameters less than half the size at fixed orbital ellipticity. 

Finally, \citet{Ludlow19} demonstrated that the common procedure of having equal numbers of DM and baryonic resolution elements leads to systematic energy exchanges between the two species. In this setup, DM particles are $\sim4$ times more massive than the baryonic elements, and through equipartition energy is transferred from the DM to the baryons. The net result is that the DM halo spuriously contracts while the stellar galaxy spuriously expands. We will not discuss this contribution to the resolution effects in this paper; it is nevertheless an important factor to consider in future work. 

Given the complexity of the physical processes in place and the interplay with their numerical implementation, and given the diversity of qualitative results claimed in past works, here we
take advantage of the range of different resolution simulations available in the IllustrisTNG suite (hereafter referred to as `TNG') to study how mass resolution affects the process of satellite stripping and the build-up of ex-situ stellar haloes. In this paper, we will focus primarily on how much specific model predictions related to the assembly of stellar haloes and intra-cluster light change with respect to resolution. We will check for the impact of spurious stripping and disruption across resolution levels and across a wide range of galaxies and host masses, by automatically accounting for the role of baryonic physics. The TNG simulation suite is well suited for this: it contains a wide range of hosts and satellites, from galaxy clusters in the TNG300 volume to dwarf galaxies in orbit around $L_{*}$ hosts in TNG50, and each volume has been simulated at three or more resolution levels, each separated by a factor of 2 (8) in space (mass). Importantly, beyond the softening choices and a parameter related to the feedback of super massive black holes \citep{Pillepich17, Pillepich17b}, all aspects and parameter choices are kept fixed in all simulations, irrespective of resolution. 
In order to evaluate the change in model predictions across resolutions, we develop a Lagrangian region matching technique that enables us to match galaxies between different resolution simulations of the same volume, and we are making these catalogues public alongside this paper. 

The resolution dependence of stripping is especially important in the context of how stellar haloes are built. These diffuse distributions of stars that inhabit DM haloes -- both the stellar haloes of isolated galaxies and the intra-cluster light of galaxy clusters -- provide vital clues about galaxy formation. First, they encode information about the merger history of galaxies and clusters \citep{Cooper13,Pillepich14b,Deason15,Merritt16,Helmi18}. They also show how much of the stellar mass in the central galaxy was assembled from accreted satellites as opposed to formed in-situ from gas that cooled in the host halo, and are also connected to the central supermassive black hole growth evolution and the central galaxy morphology \citep{Pillepich15,Deason16,Amorisco17,DSouza18,Iorio19}. Finally, stellar haloes give clues about the properties of satellites accreted at early times -- their ages, masses, and metallicities -- and can thus paint a picture of dwarf galaxies at redshifts higher than can be observed directly \citep{Fattahi19}.

Extracting surface density profiles of stellar haloes and the intra-cluster light requires deep observations of nearby galaxies, as has been recently undertaken by the Dragonfly telescope array \mbox{\citep{Abraham14}} and the LIGHTS survey \mbox{\citep{Trujillo21}}, with further improvements expected from the Vera C. Rubin Telescope. The Dragonfly facility in particular has imaged the stellar haloes of local galaxies in the Dragonfly Nearby Galaxies Survey (DNGS). An early DNGS paper detected and described the stellar haloes of eight spiral galaxies \mbox{\citep[][]{Merritt16}} -- stellar mass $[1.5,8]\times10^{10}$~$\msun$ -- out to a distance of 18~Mpc from the Milky Way. One feature common to all eight systems is that the surface brightness profiles dropped by 8~mag~arcsec$^{-2}$ between the galaxy centre and a distance of 40~kpc. These eight stellar haloes have been subsequently compared to the results of the TNG100 simulation \mbox{\citep{Merritt20}}. This comparison showed that the TNG model overestimates the surface density of stellar haloes relative to observations, especially in the halo outskirts ($>20$~kpc) where the theoretical prediction for the surface density is typically an order of magnitude higher than measured for local spirals by DNGS. \mbox{\citet{Merritt20}} explored in detail a variety of systematic uncertainties in the observational selection and in the simulations, finding that no one source of uncertainty can explain the full discrepancy between observations and simulations. A study of larger host haloes by \mbox{\citet{Ardila21}} also showed a slight excess stellar mass in the intra-cluster light of TNG galaxy groups with respect to the observations from the Hyper-Suprime Cam Subaru Strategic
Program (HSC SSP) survey, with an average of $\sim20$~per~cent excess stellar mass at 100~kpc from the host centre. However the discrepancy is smaller than the order of magnitude difference measured for the \mbox{\citet{Merritt20}} stellar haloes.  Recently, \mbox{\citet{MontenegroTaborda25}} have shown that the smooth stellar mass distribution of cluster brightest galaxies and intra-cluster light in TNG300 galaxy clusters is about twice as extended than in observations and approximately 1 mag arcsec$^{-2}$ brighter. 

With these results in mind, in this paper we consider how the stripping and disruption rates of galaxies change as a function of resolution, with a view to ascertaining the degree to which any of the resolution-related uncertainties discussed above could play a role in the aforementioned simulation-observation discrepancies. We consider the intra-cluster light of group- and cluster-mass haloes as well as $L_{*}$ stellar haloes in order to benefit from the statistics of the TNG300 volume. We are also able to take advantage of the high resolution TNG50 data, to assess lower-mass satellites. 

This paper is organised as follows. In Section~\ref{sec:methods} we present a summary of the TNG simulations, of our halo matching algorithm, and of our synthetic galaxy selection. We present our results in Section~\ref{sec:res} and our conclusions in Section~\ref{sec:summary}.

\section{Simulations and Methods}
\label{sec:methods}

\subsection{The TNG suite across resolutions}
\label{sec:sims}

The primary simulations used in this paper are the TNG100 and TNG300 collections of simulations \citep{Marinacci17,Nelson17,Naiman18,Pillepich17b,Springel17} plus the more recent TNG50 suite \mbox{\citep{Nelson19,Pillepich19}}. These together comprise the TNG project\footnote{\url{www.tng-project.org}} and all use the TNG model of galaxy formation \citep{Weinberger17,Pillepich17} and are publicly available \citep{Nelson19release}. 

The numerical model includes gas cooling, star formation, feedback from supernovae and AGN, and magnetic fields. These simulations were run with the {\sc arepo} gravity+magnetohydrodynamical code \citep{Springel10}. The TNG300 runs have a box side length of approximately 300~Mpc, TNG100 of 100~Mpc, and TNG50 of 50~Mpc. The cosmological parameters are compatible with the Planck cosmology \citep{PlanckCP13,PlanckCP15}. 

All three periodic cubes -- which we refer to hereafter as `volumes' -- were simulated at three resolution levels in DM particle mass, labelled TNGXXX-($i$), where TNG50-1 has the smallest particle mass and TNG300-3 the largest (and hence worst resolution of the series). The mass resolution of TNG300-1 is essentially the same as that of TNG100-2, and TNG100-1 has slightly worse resolution as TNG50-2: this enables us to cross-reference our results across simulations. For example, if a given quantity has converged to a given degree between TNG50-1 and TNG50-2, we have therefore shown that this quantity is converged to a similar degree in TNG100-1 where we have much better statistics. 

\begin{figure}
    \centering
    \includegraphics[width=0.45\textwidth]{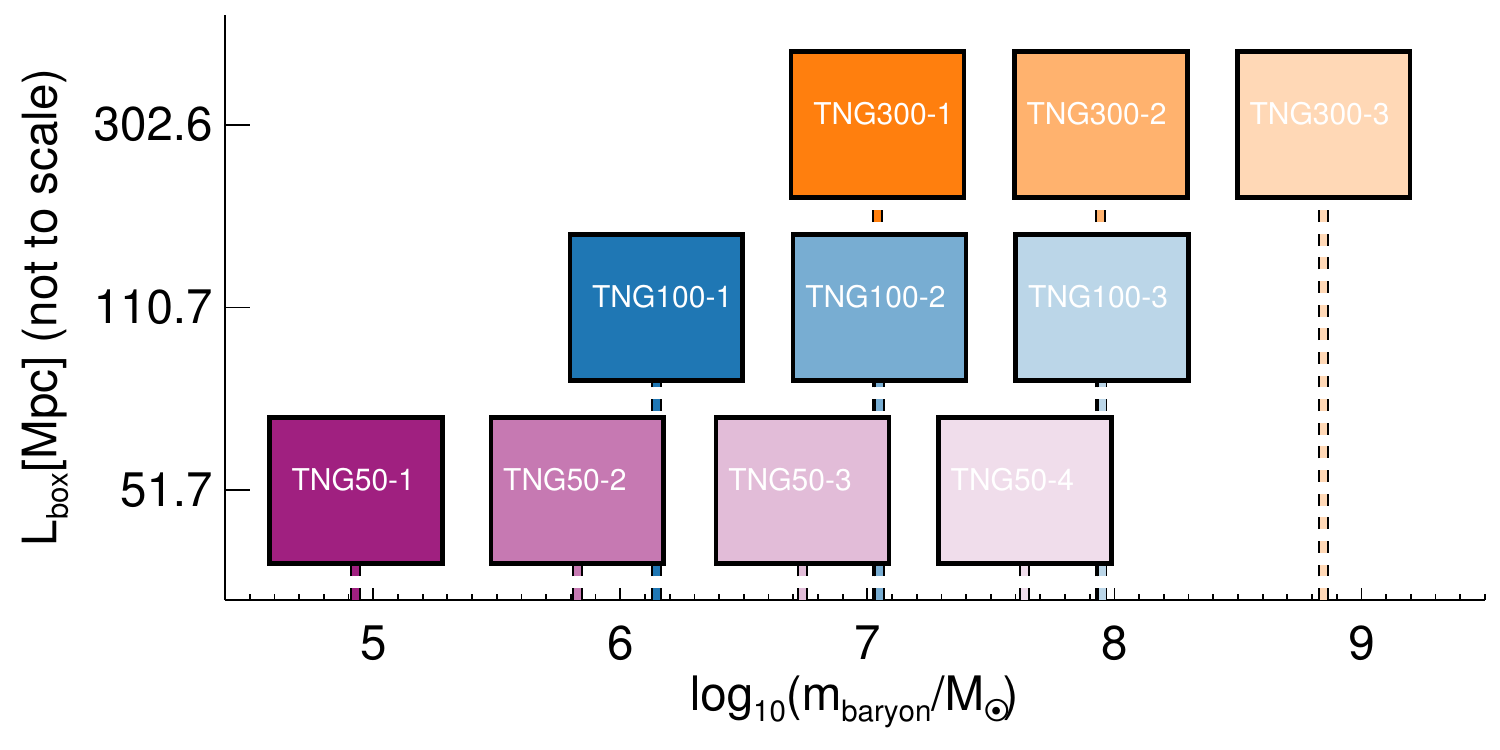}
    \caption{The baryonic particle mass and volume for each of the nine TNG simulations considered in this paper plus TNG50-4, as indicated in the box labels. The boxes are arranged such that the largest box -- TNG300 -- occupies the top row and the smallest box -- TNG50 -- occupies the bottom row. The colour of each box indicates the colour used to denote each data set throughout this paper. See Table\ref{tab1} and text for more details.}
    \label{fig:simres}
\end{figure}

For ease of comparison among simulations, in Fig.~\ref{fig:simres} we present the baryonic particle masses of each simulation box plus the colours used to denote each data set in our analysis. We provide the key details of each simulation in Table~\ref{tab1}. In addition to these nine simulations, we also include the TNG50-4 simulation with which in principle we could examine TNG50 galaxies across four resolution levels. In practice, the particle masses of TNG50-4 and TNG50-3 are very similar to those of TNG300-1 and TNG300-2 respectively, but with a much smaller volume, therefore we omit TNG50-4 from our analysis. In practice, in this paper we can quantify the effects of resolution at 9 different points for baryonic mass resolutions between $8.5\times10^{4}\msun$ (TNG50-1) and $7.0\times10^{8}\msun$ (TNG300-3).

For more details on the numerical resolution parameters of the simulations, please see papers by \citet{Nelson19, Pillepich19, Nelson19release}; we note in particular that these parameters are not altered with decreasing simulation particle mass, and therefore our study is very well suited to resolution studies. Importantly, due to the quasi-Lagrangian nature of the {\sc arepo} code, while the target mass of the gas cells is kept fixed, the spatial resolution of gas cells when they interact hydrodynamically and gravitationally is adaptive: for example, the smallest gas cells in TNG50-1 at $z=0$ span $7-8$ pc but the median cell size throughout the whole box is about 6 kpc.

\begin{table}
    \centering
    \caption{The main characteristics of each TNG simulation: simulation name, box size, $z=0$ gravitational softening length of DM and stellar particles, DM particle mass, and baryon particle mass, i.e. target mass of gas cells and stellar particles. Note that stellar particles can change mass during their lifetime due to stellar evolution; importantly, the hydrodynamical and gravitational spatial resolution of the gas cells is adaptive.}
    \begin{tabular}{cccccc}
    \hline
    Simulation &  $L$ [Mpc] & $\epsilon_\rmn{DM, stars}^{z=0}$ [kpc] & $m_\rmn{DM}$ [$\msun$] & $m_\rmn{baryon}$ [$\msun$] \\
        \hline
            TNG100-1 &  110.7 & 0.74  & $7.5\times10^{6}$ & $1.4\times10^{6}$ \\
            TNG100-2 &  110.7 & 1.48  & $6.0\times10^{7}$ & $1.1\times10^{7}$  \\
            TNG100-3 &  110.7 & 2.95  & $4.8\times10^{8}$ & $8.9\times10^{7}$\\
            \\
            TNG300-1 &  302.6 & 1.48  & $5.9\times10^{7}$ & $1.1\times10^{7}$  \\
            TNG300-2 &  302.6 & 2.95  & $4.7\times10^{8}$ & $8.8\times10^{7}$  \\
            TNG300-3 &  302.6 & 5.90  & $3.8\times10^{9}$ & $7.0\times10^{8}$ \\
            \\
            TNG50-1 &  51.7 & 0.29  & $4.5\times10^{5}$ & $8.5\times10^{4}$ \\
            TNG50-2 &  51.7 & 0.74  & $3.6\times10^{6}$ & $6.8\times10^{5}$  \\
            TNG50-3 &  51.7 & 1.48  & $2.9\times10^{7}$ & $5.4\times10^{6}$ \\
  \hline
    \end{tabular}
    \label{tab1}
\end{table}

\subsection{Halo and galaxy properties}
\label{sec:halofinders}

In all TNG simulations, DM haloes are identified using the friends-of-friends algorithm and are subsequently decomposed into gravitationally-bound structures using the {\sc subfind} code \citep{Springel01}. {\sc subfind} structures are constrained to contain at least 20 particles or cells. Each {\sc subfind} structure hosts up to one galaxy; the most massive structure in each friends-of-friends group is considered to be the host halo and its galaxy is considered to be the central galaxy; all other structures are referred to as subhaloes and to as satellite galaxies if a stellar component is present.

The goal of this paper is to characterise the evolution of several satellite galaxy properties across time, and therefore requires some nomenclature and symbol usage in order to define these properties efficiently. We describe the definitions of these properties' symbols throughout the text when they are introduced; however, for ease of reference we also list them in Table~\ref{tab:def}.

\begin{table*}
    \centering
        \caption{List of symbol definitions used in this paper.}
    \begin{tabularx}{\textwidth}{lX}
    \hline
       Label & Description \\
       \hline
         $\RTWOC$ & The radius of a central halo that encompasses a density 200 times the critical density for collapse, otherwise known as the virial radius. \\    
         
         $\MTWOC$ & Halo mass (across all species: DM, stars, gas, SMBHs) enclosed within $\RTWOC$, otherwise known as the virial mass. \\
         
         $M_\rmn{host}$ & $z=0$ value of $\MTWOC$ for a given satellite galaxy's host halo. \\
         
         $t_\rmn{lb,infall}$ & Amount of time prior to the present day that the satellite galaxy first enters its host's $\RTWOC$; also known as infall lookback time, and abbreviated as infall time. \\
         
         $M_\rmn{*-Infall}$ & Stellar mass of a satellite at the infall time, measured within a 30~kpc aperture of the satellite centre. \\
         
         $M_\rmn{dyn}$ & Total gravitationally-bound mass of a sub/halo across all mass species, also known as the dynamical mass.  \\
         
         $M_\rmn{DM}$ & DM mass of a satellite at time $t$, where $t=0$ is set at infall. Measured within a 30~kpc aperture of the satellite centre. \\
         
         $t_\rmn{50}$ & Time after infall for the value of a given mass species (DM within 30~kpc, pre-infall stellar mass within 30~kpc) to fall to 50~per~cent of its value at the infall time.\\
         
         $t_\rmn{90}$ & Time after infall for the value of a given mass species (DM within 30~kpc, pre-infall stellar mass within 30~kpc) to fall to 10~per~cent of its value at the infall time -- i.e. lose 90~per~cent of its infall mass -- and therefore approximately disrupted.\\

         $C$ & Satellite concentration, as the ratio of the species mass within 5~kpc of the satellite centre to the same species mass within 30~kpc \\
         
         $M_\rmn{*-Peak}$ & Peak stellar mass of a satellite achieved over the whole simulation time .  \\
         
         $M_\rmn{*-Strip}$ & Stellar mass of a satellite that is stripped from the satellite into its host halo.
         \\ 
         $M_\rmn{*-SB}$ & Peak value of the stellar mass after infall, including stars formed during any starbursts, and measured within a 30~kpc aperture of the satellite centre.\\ 
         
         Subscripts $A_\rmn{LX}$ & Values of some variable $A$ measured in different resolution simulations. Where a resolution is specified a number 1-3 will be used, e.g. $M_\rmn{*-Infall,1}$, $t_\rmn{50,3}$.\\
    \hline     
    \end{tabularx}
    \label{tab:def}
\end{table*}

\subsection{Matching haloes and galaxies across resolution runs}
\label{sec:matching}

One of the strengths of our analysis is that we are able to identify individual objects in different-resolution counterpart simulations of the same volume, and to show how those galaxies evolve and are processed differently, because of the different underlying numerical resolution, on a case-by-case basis. This includes matches between TNG100-1 and TNG100-(2,3), TNG50-1 and TNG50-(2,3), and between TNG300-1 and TNG300-(2,3). We caution that the properties of any two paired galaxies can be changed dramatically by stochastic effects that are not resolution dependent \citep{Genel19}, and therefore one should derive conclusions from studying the co-evolution of pairs of galaxies only across many pairs, i.e. at the population level. 

We match galaxies between counterpart simulations using the Lagrangian matching method of \citet{Lovell14,Lovell18b} . Here we present a summary of this method as applied to identify matches between galaxies at some snapshot of interest, either at $z=0$ or at higher redshift as per the science question considered. Beginning at the chosen simulation output snapshot\footnote{For this paper we use snapshots corresponding to three redshifts: $z=0,~1$ and $2$, as discussed below.}, we determine at which prior snapshot each halo attains its maximum mass, by means of the SubLink merger trees \citep{RodriguezGomez15}. We select the halo's DM particles at that maximum mass snapshot, and find the location of those particles in the simulation initial conditions, which defines the `protohalo'. We then compute the cross-potential energy of those particles with the particles of candidate halo matches in the counterpart simulation, where the cross-potential is defined as the gravitational potential energy of a protohalo due to the candidate counterpart. If the cross-potential energy is the same value as the self-potential energy of each protohalo, then we have a good match. We therefore use this cross-potential to compute a matching quality parameter, $R$, where $R=1$ corresponds to a perfect match and $R<1$ indicates an imperfect match; $R<0.5$ typically indicates the two haloes are not the same object. 

We have generated matching catalogues at $z=0$, $z=0.5$, $z=1$, and $z=2$ in order to compare the properties of galaxies both at the present time and before they fell into the host halo, and these catalogues are made publicly available with this publication. 

Note that the TNG100 box is not a subvolume of TNG300, nor TNG50 a subvolume of either larger box, therefore it is not possible to match haloes among the different TNG volumes.  
All of the matching catalogues used in this study involve matching the highest resolution, level-1 (hereafter abbreviated as `L1') version of each volume to both the level-2 (L2) and level-3 (L3) copies, and not L2 matched to L3. In some parts of the analysis below we are interested in convergence between L2 and L3, therefore, where appropriate, we select objects that appear in both the L1--L2 and L1--L3 catalogues to form an effective L2--L3 catalogue. 

\section{Results}
\label{sec:res}

In this study we are interested in two aspects of satellite stripping. The first aspect is the influence of resolution on the stripping rate of the collisionless material (DM and stars) and the disruption over multiple orbits, as discussed in Section~\ref{sec:int}. The second one is the impact of that resolution dependence on the ultimate and emergent shape and amplitude of the stellar halo profiles at low redshift. 
Although related and as we will see in the following, the resolution requirements underlying these two problems may not be the same.

We begin our results by presenting the $z=0$ subhalo and satellite abundances (Section~\ref{sec:subabund}) and the stellar density profiles (Section~\ref{sec:smprof}) in each of the simulations. Having set the context for the present day properties of the subhalo and satellite populations and of the simulated stellar haloes, we then proceed on the question of satellite stripping with two complementary angles. First, to obtain statistics in multiple-orbit and low satellite-to-halo mass ratio objects, we measure the resolution dependence for stripping and disruption rates of satellites in host haloes of mass $>10^{12}$~$\msun$ in Section~\ref{subsec:tsg}. We then apply these results to the context of intra-cluster light and stellar haloes of both group mass and $L_{*}$ galaxy halo mass haloes via the origin of the ex-situ stars, and these results are presented in Section~\ref{subsec:shpp}.

\begin{figure*}
    \centering
    \includegraphics[scale=0.50]{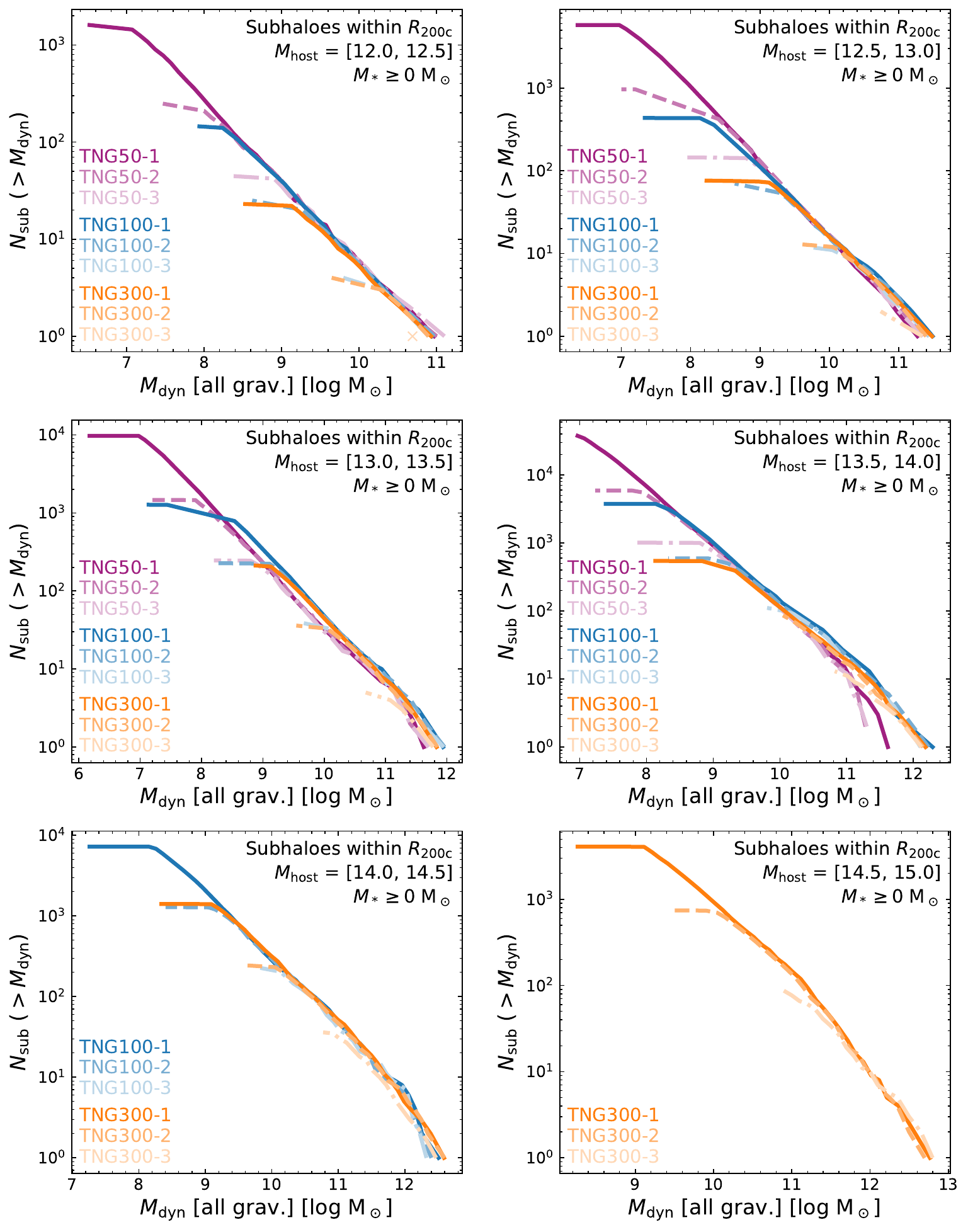}
    \caption{Subhalo mass functions for six logarithmic bins in host halo mass across the TNG simulation suite at $z=0$. The definition of mass is the total mass bound to each subhalo, here referred to as the dynamical mass, $M_\rmn{dyn}$. Level-1 resolution simulations are shown with solid lines, level-2 with faded dashed lines, and level-3 with faded dot-dashed lines. TNG100, TNG300, and TNG50 results are shown in blue, orange, and magenta, respectively. The host halo masses increase across panels, from left to right and top to bottom; the host mass values are given in each panel legend. The DM particle mass for each simulation is as follows: $[7.5,60,480]\times10^{6}$~$\msun$ (TNG100-1,2,3); $[5.9,47,380]\times10^{7}$~$\msun$ (TNG300-1,2,3); $[4.5,36,290]\times10^{5}$~$\msun$ (TNG50-1,2,3).}
    \label{fig:submf}
\end{figure*}

\subsection{Subhalo and satellite abundances in TNG across resolutions}
\label{sec:subabund}

The primary diagnostic for comparing the satellite populations across simulations is the subhalo mass function. If the number of subhaloes accreted across models and the number of surviving subhaloes to $z=0$ were to vary, in a statistical sense, this would cause changes in the amount of stellar mass available to build, e.g., the stellar halo. 

The resolution convergence of the subhalo mass functions in cosmological simulations has been quantified extensively, specifically in DM-only scenarios \citep[e.g.][]{Springel08b, BoylanKolchin09, Gao2011}. However, galaxy formation processes and the dynamics of the gas are known to modify the predicted subhalo mass function with respect to DM-only models \citep[e.g.][for Illustris and TNG]{Chua16, Engler21}. This happens for a number of reasons. On the one hand, baryons can be removed by supernova feedback and at the epoch of reionisation, which can both vary considerably among models \mbox{\citep{Sawala16b,Lovell18b}}. A second mechanism is for the baryons to modify the mass profile of both the subhaloes and the host halo, through the presence of the stellar galaxy and through steepening of the host halo through adiabatic contraction \mbox{\citep{Blumenthal86,Gnedin04}}: this can both affect the capability of the subhaloes to hold on to their material against stripping and can shorten the timescale for subhaloes to sink to the host centre and be disrupted. 

In Fig.~\ref{fig:submf} we check for variation across simulation resolutions by computing median subhalo mass functions in six host halo mass bins: thanks to the dynamical range covered by the TNG simulations (see Fig.\ref{fig:simres}), we can analyse haloes from $10^{12}$ to $10^{15} \, \msun$. For halo mass we use $\MTWOC$ (see Table~\ref{tab:def}). We adopt this definition for host halo masses throughout this paper, and apply in this context the label $M_\rmn{host}$. To label the subhaloes, we use the dynamical mass $M_\rmn{dyn}$ of the subhaloes, including all mass species, defined as the total mass that is gravitationally-bound to the subhalo according to {\sc subfind}. Here, we consider all subhaloes regardless of whether they host any star particles and that are located, at the time of inspection, within the virial radius, $\RTWOC$. In this case, the hosts are not matched between simulation realisations of the same volume. 

\begin{figure*}
    \centering
    \includegraphics[scale=0.341]{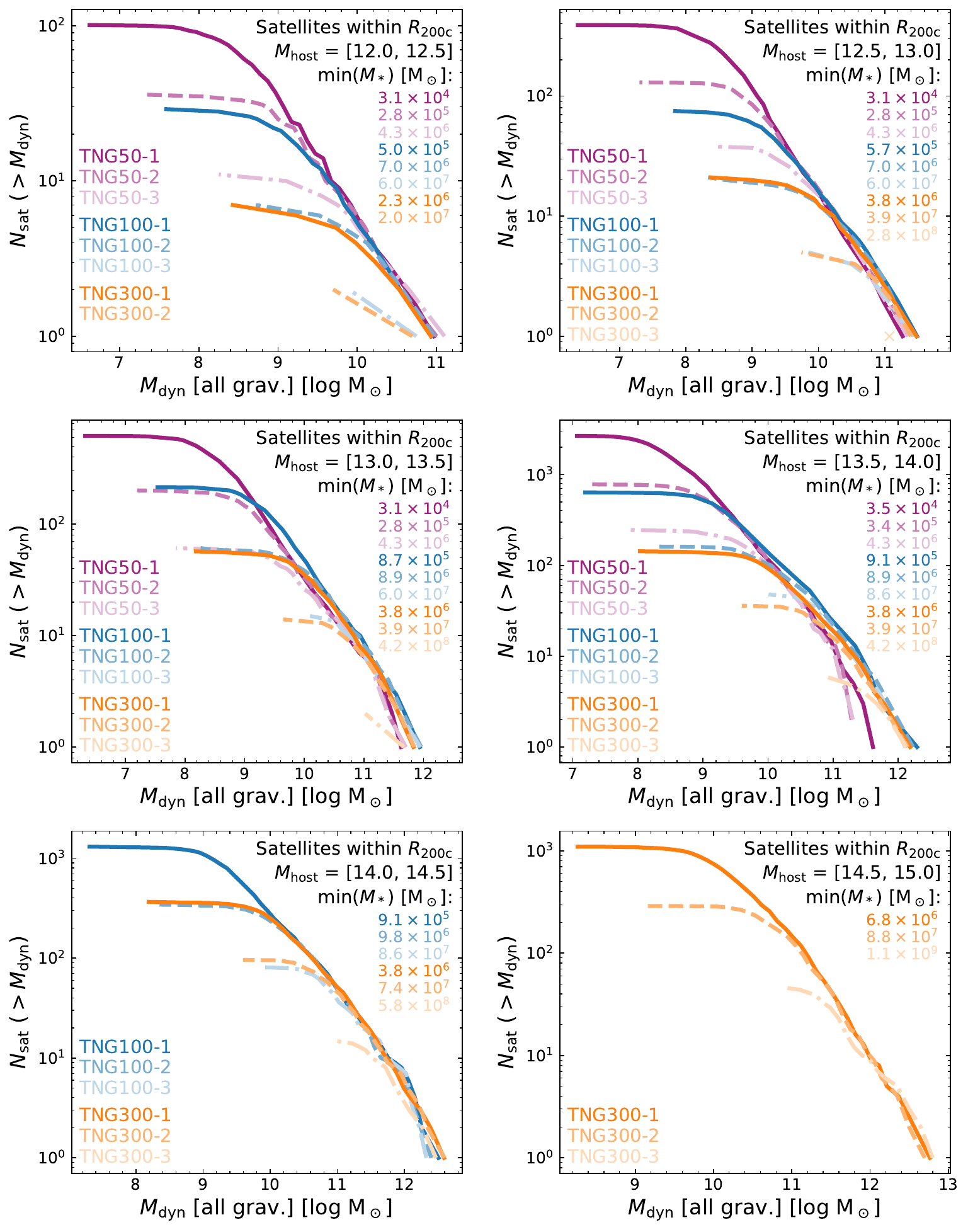}
    \includegraphics[scale=0.341]{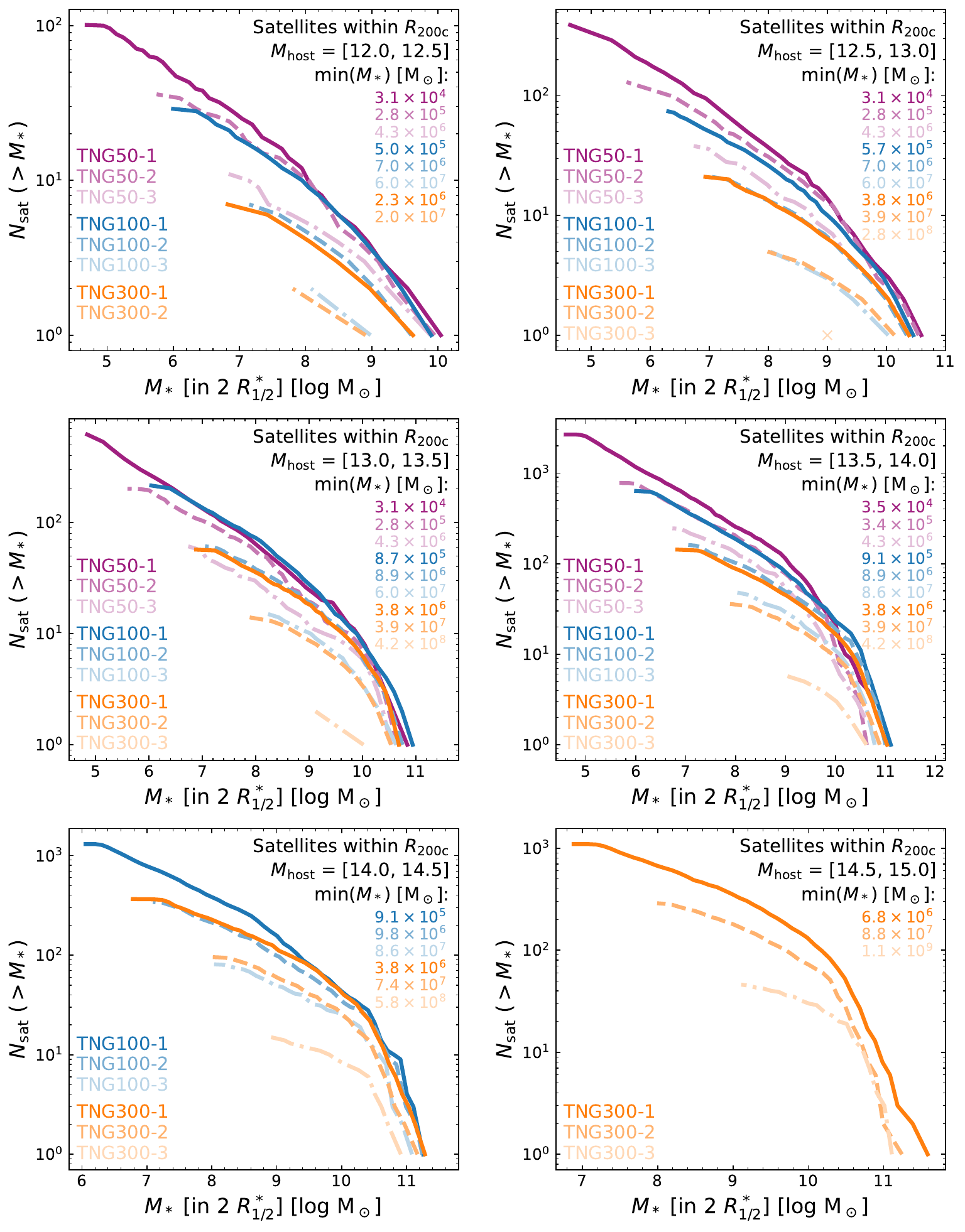}

    \caption{Satellite mass functions for hosts from $10^{12}$ to $10^{15}\,\msun$ in host halo mass across the TNG simulation suite at $z=0$. The six left-hand panels use the total mass bound to each subhalo, $M_\rmn{dyn}$, to characterize the satellites, whereas the six right-hand panels use the stellar mass within twice the stellar half mass radius. Unlike Fig.~\ref{fig:submf}, each satellite included here is required to have at least one star particle, the mass of which is given in the figure legend: i.e. we are only considering luminous satellites. Annotations and colours are as in Fig.~\ref{fig:submf}.}
    \label{fig:satmf}
\end{figure*}

As Fig.~\ref{fig:submf} shows, within each host halo mass bin and above a characteristic subhalo turnoff mass, the subhalo mass functions are consistent at roughly the per~cent level across all resolution simulations. The greatest variations are at subhalo masses $>10^{10}$~$\msun$, where stochastic differences between individual subhaloes are at their strongest, especially for the relative paucity of massive subhaloes in the TNG50 realizations of the most massive hosts, that is, those hosts in the $M_\rmn{host}=[10^{13.5},10^{14.0}]$~$\msun$ bin. A similar feature is present around subhalo masses of $10^{9.5}$~$\msun$ in the $M_\rmn{host}=[10^{13.0},10^{13.5}]$~$\msun$ bin. Only at the regime where subhaloes are resolved with $\lsim 100$ particles is there a significant deviation, with mass functions clearly suppressed at the $\gsim10$~per~cent level compared to the next highest resolution: this is expected and consistent with all previous results based on numerical cosmological simulations. We conclude that the subhalo mass function is not influenced significantly by changes in resolution for subhaloes with $>100$ particles.  

We next consider the impact of resolution on subhaloes that contain stars, for which we use the term `satellites'. These will be biased to higher DM masses than the population of all subhaloes, and will be influenced by the minimum stellar mass that can be generated at a given resolution. 
We calculate two sets of $z=0$ mass functions, first using the dynamical mass as discussed in Fig.~\ref{fig:submf} and second using the stellar mass. For this figure, we define the stellar mass of each satellite as the bound stellar mass contained within twice the half-stellar mass radius of all bound star particles. We do not expect this definition to affect our results, although we note that this measure will predict marginally lower masses with increasing concentration compared to a fixed 30~kpc aperture; we show in Section~\ref{sub:conc} that concentration does indeed increase with improved resolution. 

As shown in Fig.~\ref{fig:satmf}, there is good agreement in the dynamical mass functions of satellites with $>1000$ matter particles. Each simulation shows roughly the same power law mass function in each host mass bin, down to a low resolution turn off where the fraction of subhaloes that host luminous subhaloes becomes resolution dependent: at the turnover the number of satellites is suppressed by up to a factor of two relative to the higher resolution simulations. In general there is good agreement between equal resolution simulations of different volumes, e.g. TNG100-2 and TNG300-1. On the other hand, the TNG50-2 and TNG50-3 runs have somewhat better resolution, and both return more satellites, than TNG100-1 and TNG100-2, as expected: see again Fig.~\ref{fig:simres}.

The results of the stellar mass functions are more complex. It has been shown that, within the TNG model, an improved resolution implies more star formation and less effective feedback, and hence ultimately larger galaxy stellar masses at fixed halo mass \mbox{\citep{Pillepich17,Pillepich17b, Engler21, Engler21b}}. This result is also reproduced here: lower particle resolution masses correspond to systematically larger numbers of galaxies at fixed stellar mass. For example, at $M_{*}\sim10^{8}$~$\msun$ the number of satellites increases by a factor of $\sim2$ between TNG300-1 and TNG100-1, and by $\sim50$~per~cent from TNG100-1 to TNG50-1. On the other hand, there is good agreement between simulations of equal resolution in different volumes. 

\begin{figure*}
    \centering
    \includegraphics[scale=0.43]{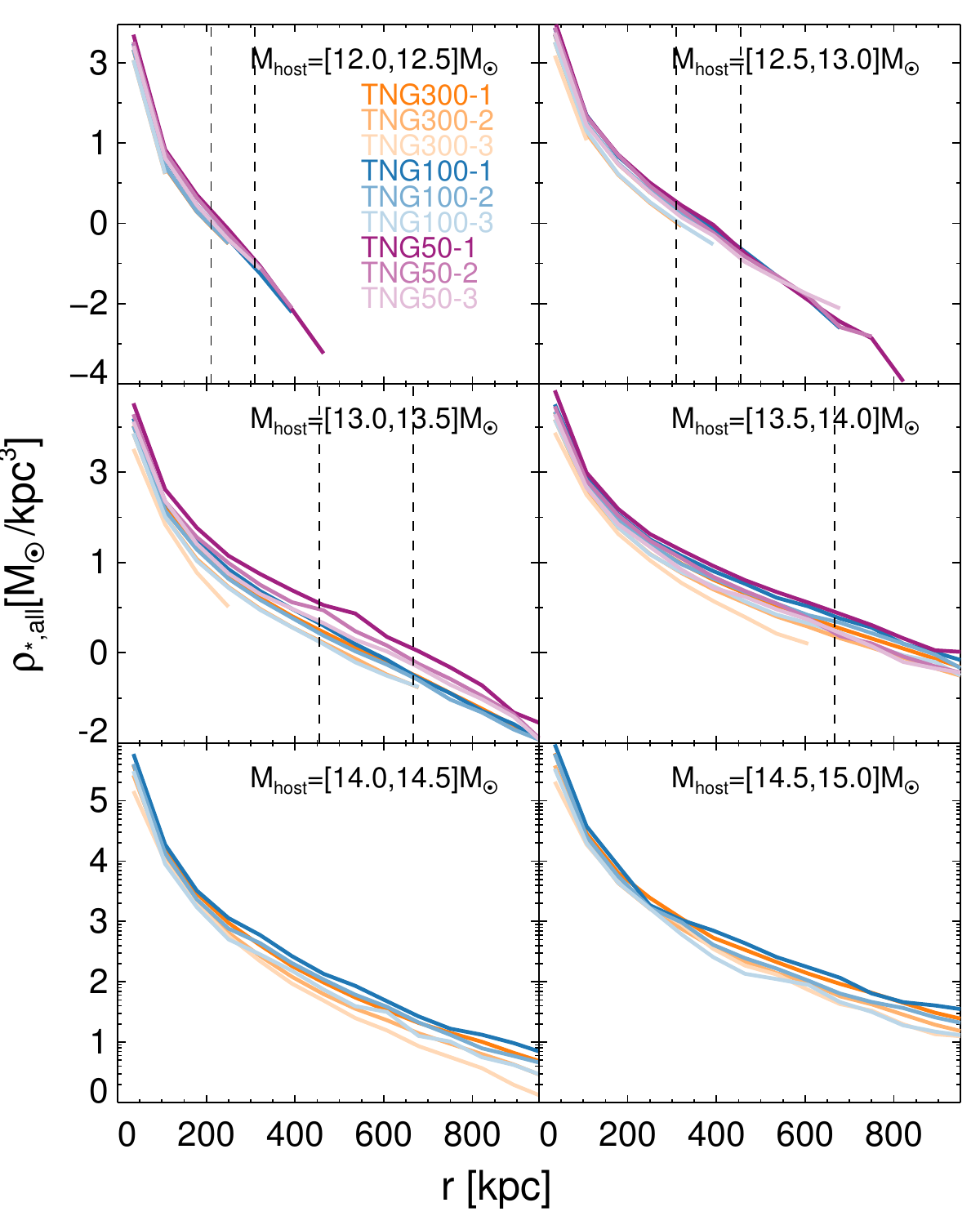}
      \includegraphics[scale=0.43]{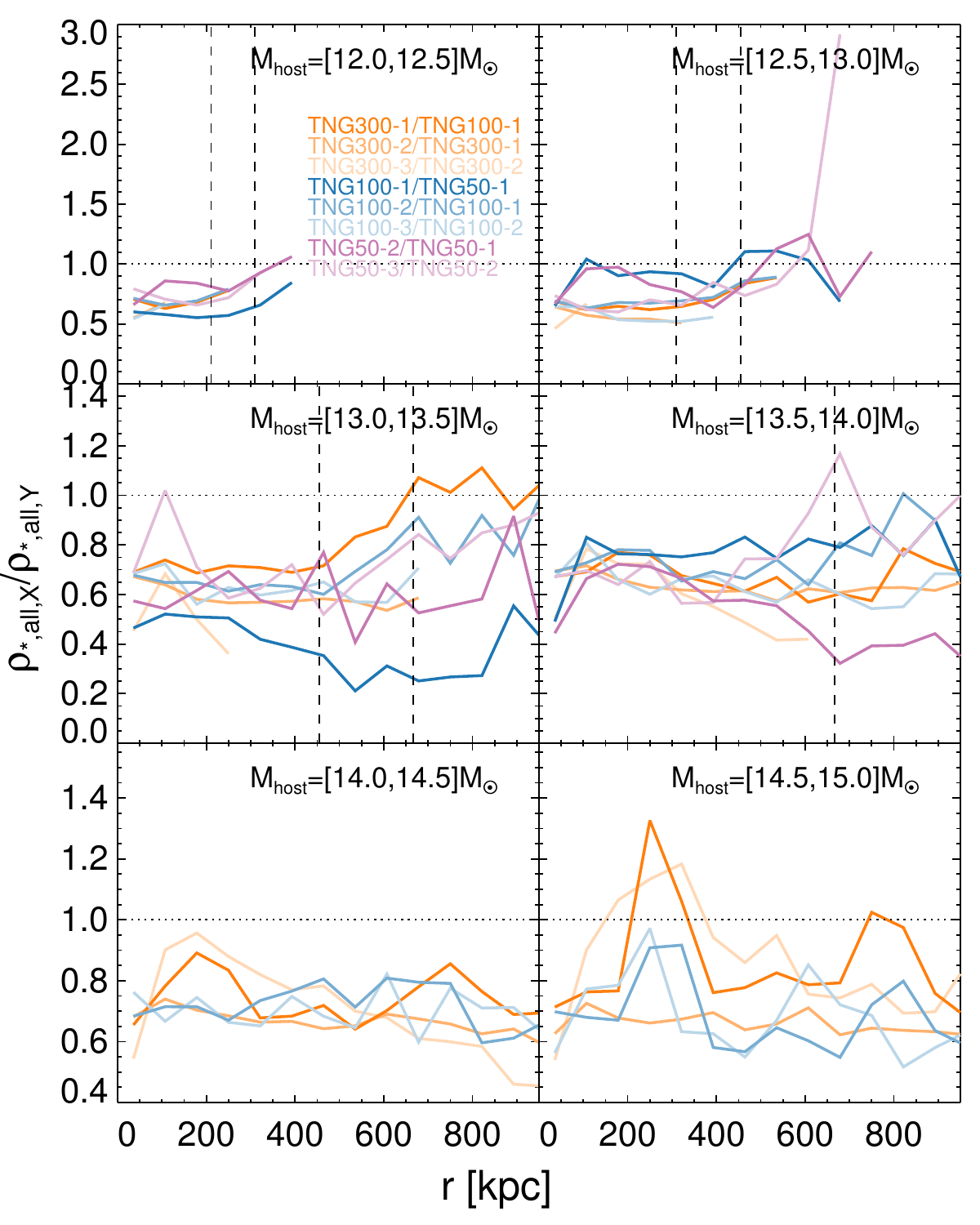}
    \caption{The stellar density profiles of central galaxies (left; median across galaxies) and the ratio of these profiles between pairs of simulations one step in resolution apart (right), across all TNG runs at $z=0$ (see colours). Each panel corresponds to a range of host halo $\MTWOC$ as indicated, shown as a pair of dashed vertical lines, if within the plotting range. On the right, for example, for the TNG300-1 curve we quantify the ratio to TNG100-1 and for TNG100-1 we compute the ratio to TNG50-1 }
    \label{fig:sh}
\end{figure*}

The gap between TNG300-1 and TNG100-1 is bigger than the gap between TNG100-1 and TNG50-1 and this is a consistent feature across host mass bins: given that the satellite stellar mass functions are progressing towards convergence with better resolution, poorer resolution therefore implies artificially suppressed satellite mass functions. Importantly, by focusing on hosts of Milky Way-like mass, we have shown in \cite{Engler21b} that this mostly appears to be related to the reduced stellar masses in subhaloes of a given dynamical mass (shifts in the x-axis) instead of the enhanced disruption of subhaloes or satellites at progressively poorer resolution (shifts in the y-axis): see Fig.~A2 of \cite{Engler21b}. In conclusion, we find that changes in resolution do not necessarily alter the rate at which hosts strip and destroy subhaloes (luminous or not). However, better resolution may imply a larger amount of (accreted i.e. ex-situ) stellar mass available to build  e.g. the stellar haloes of the host galaxies; we study this next. 

\subsection{Stellar density profiles in TNG across resolutions}
\label{sec:smprof}

Do different levels of numerical resolution return different amounts of stars at large distances from galaxies, in the so-called stellar halo or intra-cluster light?
Here we measure the $z=0$ total stellar density radial profiles, i.e. including both ex-situ and in-situ stars and denoted $\rho_\rmn{*,all}$, around the central galaxies that reside in host haloes in the mass range $\MTWOC=[10^{12},10^{15}]$~$\msun$. We only account for stellar particles that are gravitationally-bound to the central galaxy.

In Fig.~\ref{fig:sh}, left panels, we first compute stellar density profiles (3D) of each host that falls into a given mass range and then show the median profile in each mass bin including hosts with at least ten star particles. In the right panels of Fig.~\ref{fig:sh}, instead, we show the ratio of each of these curves to the median stellar density curve of a simulation that is one step up in resolution, i.e. uses a factor of $2^3$ more particles. The lower resolution counterpart is denoted $\rho_\rmn{*,all,LX}$ and the higher resolution $\rho_\rmn{*,all,LY}$. For example, for TNG300-1, we quantify the ratio to TNG100-1 and for TNG100-1 we compute the ratio to TNG50-1; TNG50-1 is the highest resolution available simulation and therefore does not have a ratio curve.

In most circumstances there is a clear increase in stellar density at fixed radius with resolution. Increasing the number of resolution elements by a factor of $2^3$ typically adds an extra 30~per~cent in stellar mass density at radii smaller than $\RTWOC$ (i.e. about 0.1 dex), but the increase can be as little as 10~per~cent or as much as 50~per~cent. Therefore, the increase with resolution is not systematic; in addition, the resolution effects are quite flat with radius -- we shall not comment too quantitatively outside of the virial radii because these low density regions are affected by poor statistics in the number of available star particles. We have performed this analysis for the ex-situ component alone, and found that there is no qualitative difference in our results.

We conclude that increasing the number of resolution elements, i.e. decreasing the resolution element mass by a factor of 8, increases the stellar densities by an average of $\sim$30~per~cent at all radii. Therefore, there is no clear evidence, based on the range of particle masses tested here, that increasing the resolution further would achieve convergence to any measurable standard. 

\subsection{Tracking and simulating stripped galaxies across TNG runs}
\label{subsec:tsg}

The analyses in the previous sections seem to indicate that the effects of numerical resolution on the rate of stripping and of disruption of satellite galaxies do not significantly alter the satellite mass functions and stellar density profiles that ultimately emerge at $z=0$ in TNG groups and clusters. In the following, we explicitly and quantitatively check this.

We first introduce our sample selection of satellites, show examples of how stripping and disruption rates of the DM and stars vary with resolution on a case by case basis, and then demonstrate the systematic impact of resolution on the relevant timescales. We end by comparing and contrasting our results with those of \mbox{\citet{vdBosch18}} to determine whether spurious disruption has an impact on stripping rates in the case of large-scale, realistic, galaxy formation simulations. 

\subsubsection{Sample selection and methodology}
\label{sec:stripping_methods}

Our synthetic satellite galaxy sample is constituted of satellites that fall into proto-groups/proto-clusters at $z\sim2$; we choose this timescale to provide 10~Gyr for satellites to be potentially stripped or disrupted. We assemble this sample as follows.
 
\begin{itemize}
    \item We identify host haloes with $\MTWOC>10^{12}\msun$ at L1 resolution that have matches at L2 and L3 at $z=0$ using our $z=0$ matching catalogue (see Section~\ref{sec:matching}).
    \item We track the location of these host haloes back to $z=2$.
    \item We identify L1 galaxies that pass within these hosts' $\RTWOC$ for the first time at lookback times $t_\rmn{lb}=[9.5,10.5]$~Gyr. This time is our definition of `infall time' (see Table~\ref{tab:def}).
    \item From this selection of satellite galaxies, we retain those L1 galaxies that have matches in the L2 simulation in the $z=2$ Lagrangian matching catalogue and, for the remainder of the analysis, we also consider these low-resolution counterparts, in both the L2 and L3 (if any) runs. We require the latter to contain at least 200 DM particles at $z=2$. 
    \item Importantly, we allow all three matched galaxies to be stripped below the resolution limit of the simulation before $z=0$ i.e. below the minimum mass i.e. minimum number of resolution elements for a (sub)halo to be stored in the catalogues (see Section~\ref{sec:halofinders}).
\end{itemize}

This process provides a sample of tens to thousands of simulated galaxies per simulation pair. We have checked, although do not show, that the infall time distributions have largely the same shape in all three simulated volumes, showing that we are not affected strongly by resolution- and selection-dependent variations across our bin of $9.5 - 10.5 $ billion years in lookback time. On the other hand, necessarily, this sample of galaxies does not represent the whole (sub)halo or satellite population in the simulations, as it is biased towards higher galaxy stellar mass at infall ($\gsim10^{7}$~$\msun$ for TNG50,$\gsim10^{8}$~$\msun$ for TNG100, and $\gsim10^{9}$~$\msun$ for TNG300) and towards a flatter number of galaxies per host halo mass than a simpler volume-limited selection would imply. The comparisons between volumes of same-resolution, i.e. TNG100-2 versus TNG300-1 and TNG50-2 versus TNG100-1, are not exact as the higher resolution volume will be relatively biased towards lower mass haloes not resolved in the lower resolution L2 iteration. We are also unable to track the disruption rates of galaxies unresolved in any L1 simulation, although we expect the ultimate impact of these objects on the intra-cluster light to be small. Nevertheless, this sample is well suited to address specific questions and quantification of stripping and survival across long periods of satellite-host interaction. 

In the following, we bin our satellite galaxy sample in a 2-dimensional grid of $z=0$ host total mass and galaxy stellar mass at infall, both defined using the values at L1 resolution. We use the same sets of bins for all runs of the TNG simulations, with four logarithmic bins for the former, in the range $[10^{7},10^{12}]$~$\msun$, and six $M_\rmn{host}$ bins in the range $[10^{12},10^{15}]$~$\msun$.

It is to be expected that the times and rates over which any given satellite galaxy is stripped are influenced by a wide range of parameters and processes: orbital parameters, mass and concentration of the satellite at infall, and the evolution and properties of the central galaxy over the subsequent 10~Gyr. In order to distinguish the aggregate behaviour of the satellite population from variations between matched pairs of individual satellites, we take the following approach. For both the DM mass and the infall stellar mass we first present the detailed mass evolution of a series of example satellites from our samples (Section~\ref{sec:sde}); we then collapse the mass evolution tracks of the whole sample into the time taken to strip the satellites to some percentage of their infall mass (Sections~\ref{sec:pst_dm} and ~\ref{sec:pst_stars}).

\begin{figure*}
    \centering
     \includegraphics[scale=0.75]{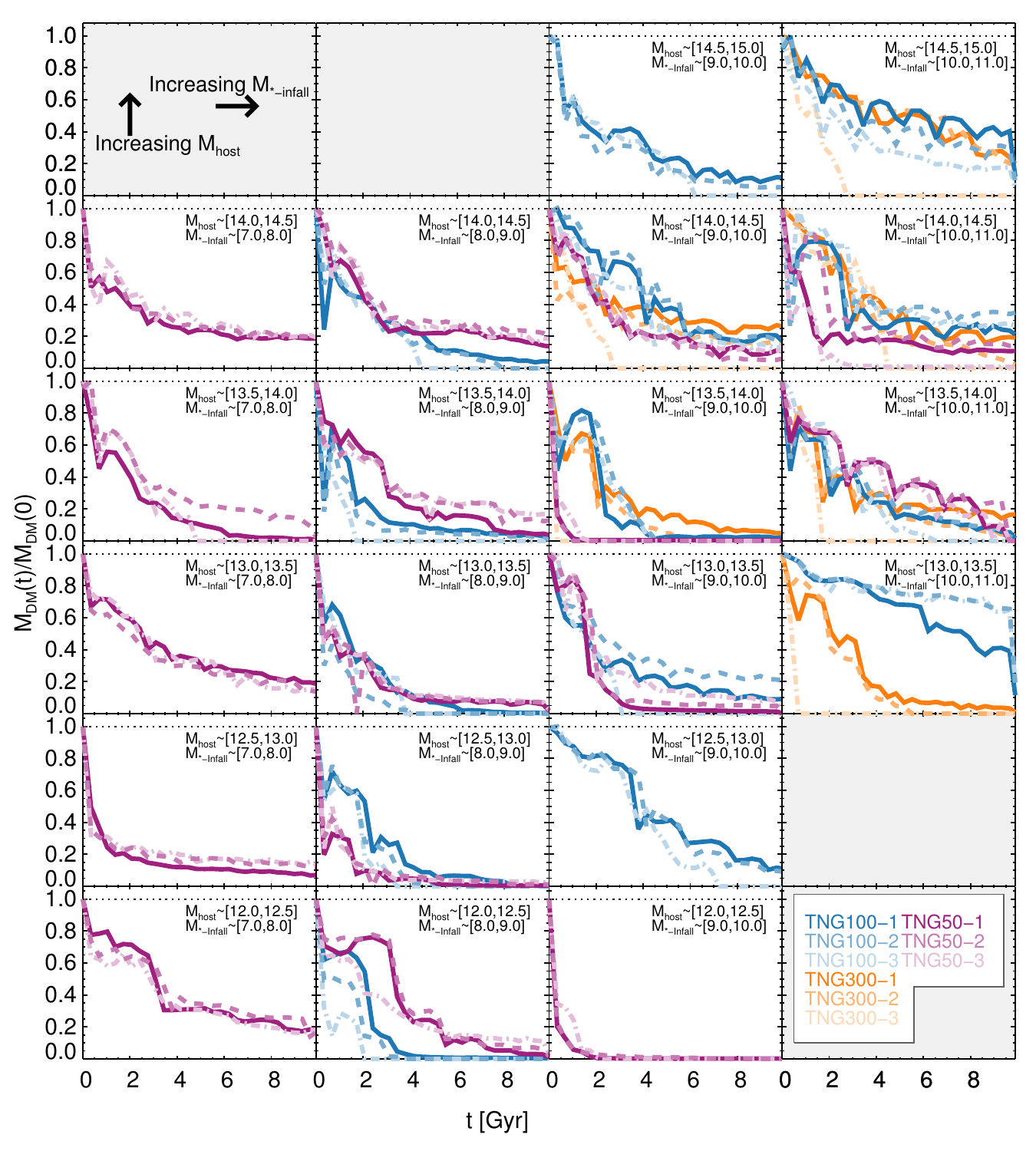}
    \caption{The evolution with time of the DM mass in randomly-selected TNG example satellites, starting at infall. Columns (rows) denote bins in satellite stellar mass at infall (host halo mass at $z=0$). Here we show the evolution of satellite galaxies matched across runs at different resolution levels and selected in the highest one (L1, i.e. selected in TNG50-1, TNG100-1 and TNG300-1 as described in Section~\ref{sec:stripping_methods}). Results from the L1, L2, and L3 runs are shown as solid, dashed and dot-dashed curves, respectively, with different colours denoting different simulated volumes. The smaller the resolution effects on the stripping of DM from satellites, the smaller is the distance among curves of the same colour.}
    \label{fig:IEDM}
\end{figure*}

\begin{figure*}
    \centering
     \includegraphics[scale=0.75]{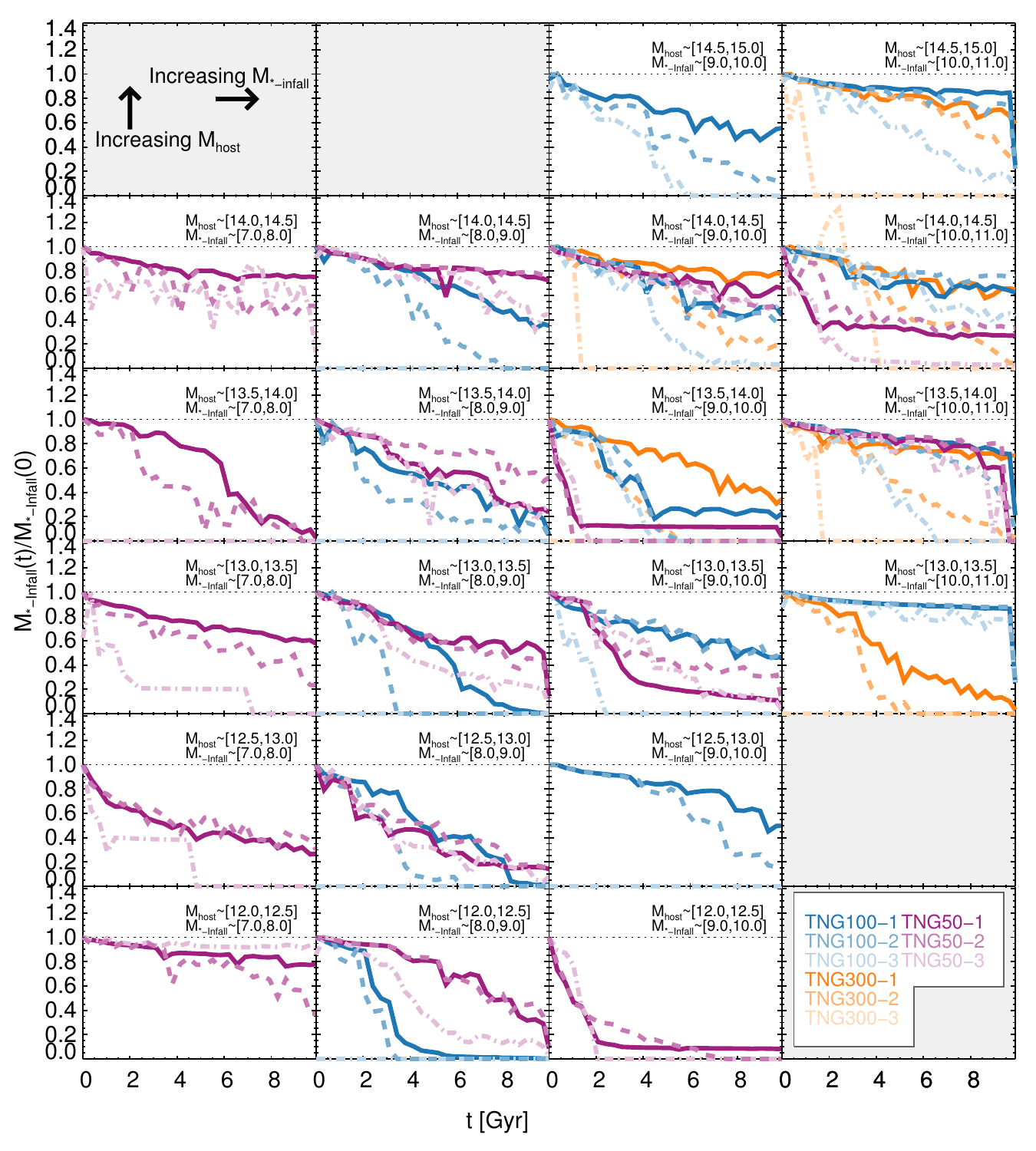}
    \caption{The evolution of the stellar mass that formed prior to infall, for the same example satellites presented in Fig.~\ref{fig:IEDM}. Columns denote bins in $M_\rmn{*-Infall}$, and rows denote bins in $M_\rmn{host}$. Median relations are shown as thick lines. The correspondence between colour and data set is given in the bottom right-hand panels. L1 satellites are shown in solid lines, L2 as dashed lines, and L3 as dot-dashed lines.}
    \label{fig:IEStarmP}
\end{figure*}

\subsubsection{Stripping of DM and stars and disruption examples}
\label{sec:sde}

Through the selection procedure presented above, we select galaxies at random as examples to compare stripping histories. We select one galaxy per host mass--stellar mass combination where available; in this manner we identify 6, 14, and 16 satellites in the TNG300, TNG100, and TNG50 samples, respectively, where TNG300 lacks galaxies with stellar masses below $\sim10^{9}$~$\msun$. For these satellites we compute the ratio in the DM and stellar mass relative to that at infall as a function of time and present the results in Fig.~\ref{fig:IEDM} for DM and in Fig.~\ref{fig:IEStarmP} for the stellar content. In both cases, we compare the rate of mass loss (most certainly via tidal stripping) of these satellites to that of the analogue ones in the lower-resolution simulations: smaller effects of numerical resolution manifest themselves in smaller differences between curves of the same colour (i.e. volume).  

At all times, the DM mass is measured within 30~kpc and is labelled $M_\rmn{DM}$. Fig.~\ref{fig:IEDM} shows that there is some variety in the rate of resolution dependence among volumes and even for individual satellites. Overall, given the better (worse) numerical resolution of the TNG50 (TNG300) runs, it is unsurprising that the most consistent agreement between resolution levels occurs for the TNG50 series (magenta), whereas the least agreement can be seen for the TNG300 runs (orange). Because of their very low resolution compared to the other runs (see Table~\ref{tab1} and Fig.~\ref{fig:simres}), all of the featured TNG300-3 galaxies are completely disrupted below the resolution limit of 20 particles within 4~Gyr of infall, even at times when the counterpart galaxies have lost less than 40~per~cent of their infall mass. Also, a few of the TNG100-3 galaxies are disrupted within 5~Gyr of accretion whereas their TNG100-1 versions survive through to the end of the simulation. On the other hand, however, for most of these examples the differences across satellites of the TNG50 and TNG100 series, and even between TNG300-1 and TNG300-2, are smaller than 10-20 percentage points; this signals a reasonable agreement, albeit with lower-resolution satellites being typically stripped more than their higher-resolution counterparts. We do notice some cases (e.g. TNG100-1 in the $M_\rmn{host}=[10^{13},10^{13.5}]$~$\msun$--$M_\rmn{*-Infall}=[10^{10},10^{11}]$~$\msun$ bin) whereby a TNG100-1 satellite deviates to higher stripping rates than its lower-resolution counterparts -- we attribute this to effects of stochasticity and present a brief discussion of the changes in orbits between resolutions in Appendix~\ref{app:orb}. Furthermore, in some satellites for which the stripping rate is less severe, there is an oscillation feature by which the DM decreases and then increases again over a period of $\sim1$~Gyr -- we attribute this behaviour to satellite mass being inaccurately associated to the host halo when the satellite is close to pericentre \citep{RodriguezGomez15,Springel21}. 
Amid the complexity due to stochasticity, the challenges of simulating satellites' orbits over multiple pericentric passages, and the possible inaccuracies of the adopted halo finders \mbox{\citep[see][for a comparison of subhalo finders]{Onions12}}, Fig.~\ref{fig:IEDM} highlights that convergence in the stripping of DM does depend on the absolute value of the numerical resolution under scrutiny but overall the systematic trends with improved resolution are small, if not negligible. Also, whereas there may be a strong relationship between numerical resolution and satellite survival (see also subsequent Sections), a weaker relationship is in place according to the TNG simulations between resolution and the stripping rate.  

Compared to the DM, tracking the stripping of stellar mass is more challenging. Increasing the mass resolution -- i.e. using a larger number of lower mass gas cells -- increases the star formation rate of galaxies, therefore increasing the stellar mass and potentially also increasing the size. Additionally, at the point of infall some satellites undergo further star formation, sometimes including an intense starburst up until the first pericentric passage, which we discuss in Appendix~\ref{app:sb}, and it is difficult to define a stripping rate when the stellar mass is increasing. We therefore restrict our analysis to the evolution in the mass in star particles formed before infall, denoted $M_\rmn{*-Infall}(t)$. We follow the stellar mass for these stellar populations in the same example satellites considered in Fig.~\ref{fig:IEDM}, and present the results in Fig.~\ref{fig:IEStarmP}. Note that sometimes the low resolution counterparts do not contain enough star particles to make a measurement: in these cases the curve rests at zero across the entire panel. 

The stripping rates for stellar mass are typically smaller than for DM: namely, at the best resolution level (L1) galaxies retain more than 60~per~cent of their stellar mass over the full time period whereas all lose at least half of the DM. The rate of stripping is frequently stronger for the lower resolutions, and in some cases more so than for the DM although more variable. In all but five satellites the stripping rate correlates with resolution, with higher resolution simulations returning less stripping than lower resolution ones, although the degree of change in stripping varies considerably. In some cases the change in stellar mass follows closely the change in DM mass (TNG50 galaxies in the $M_\rmn{host}=[10^{13.0},10^{13.5}]$~$\msun$-$M_\rmn{*-Infall}=[10^{7},10^{8}]$~$\msun$ bin) but not in others (the TNG100-1 galaxy at $M_\rmn{host}=[10^{13.0},10^{13.5}] $~$\msun$-$M_\rmn{*-Infall}=[10^{10},10^{11}]$~$\msun$ bin experiences the same stellar stripping rate as TNG100-2, even while its DM stripping rate is more severe). There is still evidence for oscillations in the stripping rate around pericentre but these are less pronounced than is the case for the DM, likely because the stellar density profile is more centrally concentrated. We conclude that the disruption rates and systematics affecting stellar mass loss are very similar to the DM results, except that the stellar mass stripping rates are lower and more strongly influenced by resolution changes.

\subsubsection{Stripping times of DM across populations of satellites}
\label{sec:pst_dm}

We extend the results obtained thus far to our full satellite galaxy sample. For all the satellites that we can match across resolution levels (see Section~\ref{sec:stripping_methods}), we analyse the stripping rate by computing the time taken for the DM measured within 30~kpc to drop below 50~per~cent of the infall value, and label this timescale $t_\rmn{50}$. We do not attempt to correct for drops in mass due to the inaccurate attribution of satellite mass to the host, and therefore these results can be thought of as a strictly lower limit. Our hypothesis based on Fig.~\ref{fig:IEDM} is that the stripping of DM is largely independent of resolution, whereas disruption does depend on resolution. We use $t_\rmn{50}$ as a proxy for the stripping timescale, and later use the time to strip away 90~per~cent of the infall DM mass, as a rough proxy for the disruption rate, and therefore label this timescale $t_\rmn{90}$. We hypothesise that $t_\rmn{50}$ does not vary systematically with resolution but that $t_\rmn{90}$ does.

\begin{figure*}
    \centering
     \includegraphics[scale=0.7]{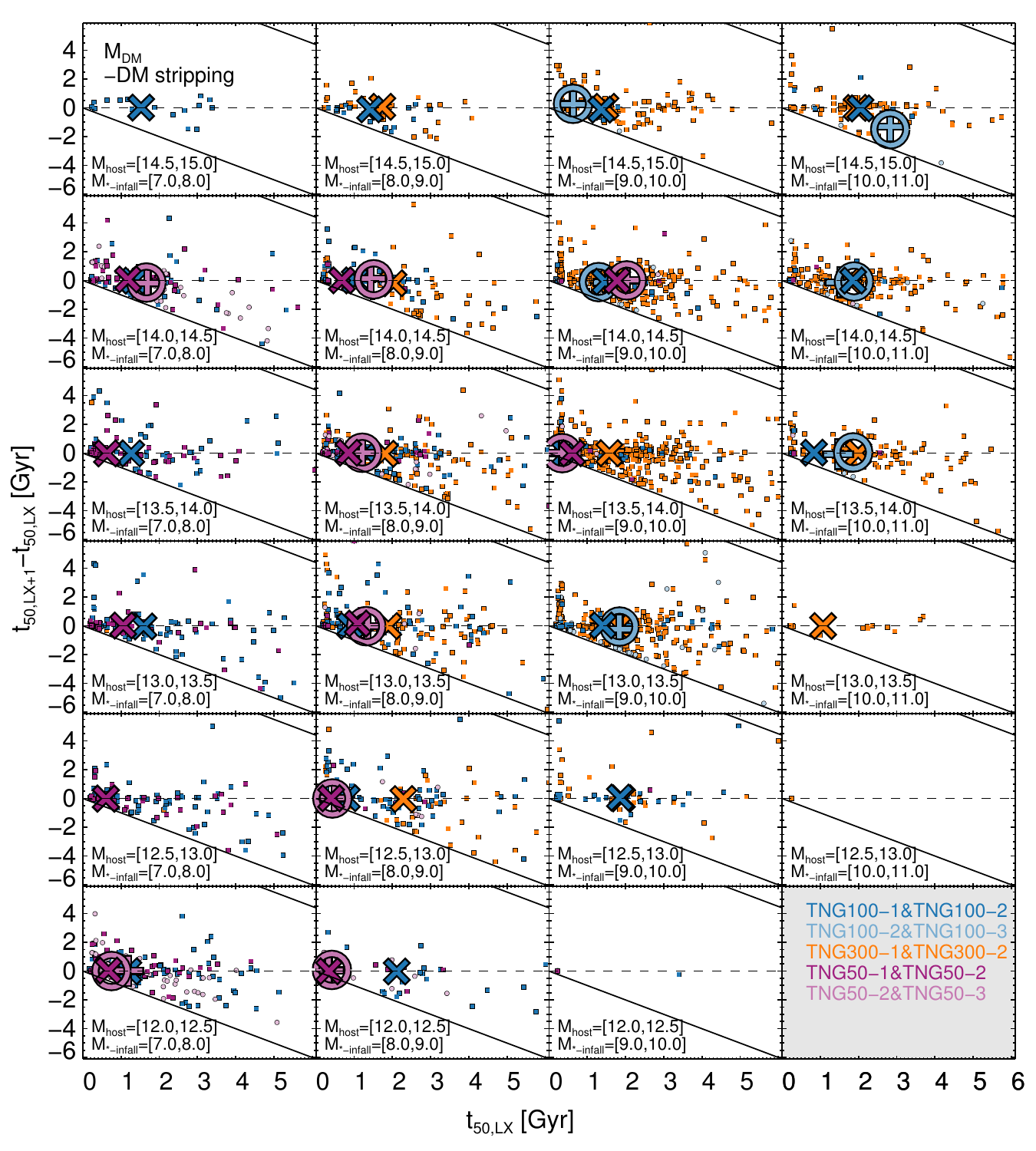}
        \caption{The time after infall at which galaxies are first stripped to 50~per~cent of their infall DM mass as measured within 30~kpc, $t_\rmn{50}$. On the $y$-axis we show the difference in stripping time between high- and low-resolution copies of satellites: L3 minus L2 and L2 minus L1. On the $x$-axis we show the $t_\rmn{50}$ of the higher resolution counterpart. If the number of L3--to--L2 matches in a panel is at least half the number of L2--to--L1 matches, we plot data for both sets of pairs; otherwise we only plot the L2--to--L1 data. Points below the 1:1 line (i.e. the dashed line at ratio equal 0) denote satellite pairs for which the low resolution counterpart is stripped earlier than the higher resolution version, and the opposite is true for points above the 1:1 line. The median data points are shown as large plus symbols (L3 versus L2) and large crosses (L2 versus L1) when binned by the L1 $M_\rmn{*-Infall}$; we also include L3 versus L2 medians when binned by L2 $M_\rmn{*-Infall}$ as empty circles. The diagonal black lines delineate parts of the plotting region that cannot include data, because either $t_\rmn{50}$ of the low resolution counterpart would occur prior to infall or later than 10.5~Gyr, i.e. at a time later than $z=0$. Each panel corresponds to a different halo mass / satellite infall stellar mass bin combination, which are binned by the $M_\rmn{host}$ and $M_\rmn{*-Infall}$ of the L1 counterpart. The relationship between symbol colour and data set combination is given in the bottom right panel figure legend. On average, and barring the worst resolution level at hand (TNG300-3), the same satellite across resolution levels is stripped of its DM mass at similar rates. However, poorer resolution tends to strip faster those satellites that take longer to be stripped (see markers below the zero line and with larger times after infall to be 50 per cent stripped).}
               \label{fig:STDM1}
\end{figure*}

In order to establish how $t_\rmn{50}$ changes with resolution we first compute the difference in L3 and L2 counterpart $t_\rmn{50}$ values where available, and then between L2 and L1 counterpart $t_\rmn{50}$ irrespective of whether an L3 counterpart exists. We plot the results as a function of the higher resolution counterpart $t_\rmn{50}$ in Fig.~\ref{fig:STDM1}, and show results in bins of both host mass (total mass, $M_\rmn{host}$) and satellite mass (stellar mass at infall, $M_\rmn{*-Infall}$; see Section~\ref{sec:stripping_methods}). As $M_\rmn{*-Infall}$ is lowered by poorer resolution, some L3 galaxies will no longer be resolved. Therefore, in $M_\rmn{host}$--$M_\rmn{*-Infall}$ bins for which fewer than half of the L1--L2 pairs have an L3 counterpart we only plot the L1--L2 data; otherwise we plot both L1--L2 and L2--L3 results. This condition is not satisfied for TNG300-3 in any of our panels, therefore no TNG300-3 satellites are included in the remainder of this subsection.

Finally, we need to consider our choice of infall stellar mass with which to bin the satellites. In Fig.~\ref{fig:IEDM} and \ref{fig:IEStarmP} our interest was in comparing the same object at three different resolution levels, but it is less clear whether this is the correct approach to examine the population given that the stellar mass of the L2 galaxies will be systematically lower than for the L1 counterparts. For example, if we wish to compare the difference in stripping times between TNG300-1 and TNG300-2 on the one hand versus TNG100-2 and TNG100-3 on the other, one should instead bin the TNG50-2 versus TNG50-3 data by the TNG50-2 infall stellar mass. Therefore, in addition to plotting the data and medians from binning L2 versus L3 results by L1 stellar mass, we also plot medians for L2 versus L3 binned by L2 stellar mass as open circles. We note, however, that the typical increase in stellar mass with each $2^3$ step in particle mass is up to a factor of 2, compared to a whole factor of 10 in the width of our $M_\rmn{*-Infall}$ bins. 
 
 Fig. ~\ref{fig:STDM1} shows that, for all of the simulation combinations, the median difference between low resolution and high resolution counterparts is zero to within a few per~cent, and therefore we can conclude that even the TNG300-2 and TNG100-3 simulations are converged in the DM stripping $t_\rmn{50}$, at least in the aggregate. However, although matched pairs of satellites are converged, there is a clear trend in most panels -- as indicated by the distribution median symbols -- for satellites selected at higher resolution to be stripped more quickly than those selected at lower resolution. For example, in the $M_\rmn{*-Infall}=[10^{8},10^{9}]$~$\msun$ panels the TNG300-1-selected satellites (purple crosses) attain $t_\rmn{50}$ on average at least 1~Gyr later than the TNG50-1-selected satellites (orange crosses). The combination of the fixed stellar mass bin and the increased star formation efficiency towards higher resolution in principle suggests that TNG300-1 subhalo dynamical masses will be higher than for TNG50-1, therefore it could take a longer time to strip to 50~per~cent of the infall mass provided the merger time under dynamical friction is small. However, binning the L2 versus L3 data by the L2 $M_\rmn{*-Infall}$ values (plus signs) still leads to results more similar to the L1 binning in the same volume (circles) than to other volume pairs run with the same resolution, therefore the change in stellar mass--halo mass relation may be smaller than differences due to the assembly histories of the host haloes. The tails of the distributions tend toward haloes that are stripped more quickly at low resolution than high resolution, although in some bins there exists a population of satellites that are stripped almost immediately after infall at high resolution while persisting for longer at lower resolution.  

\begin{figure*}
    \includegraphics[scale=0.7]{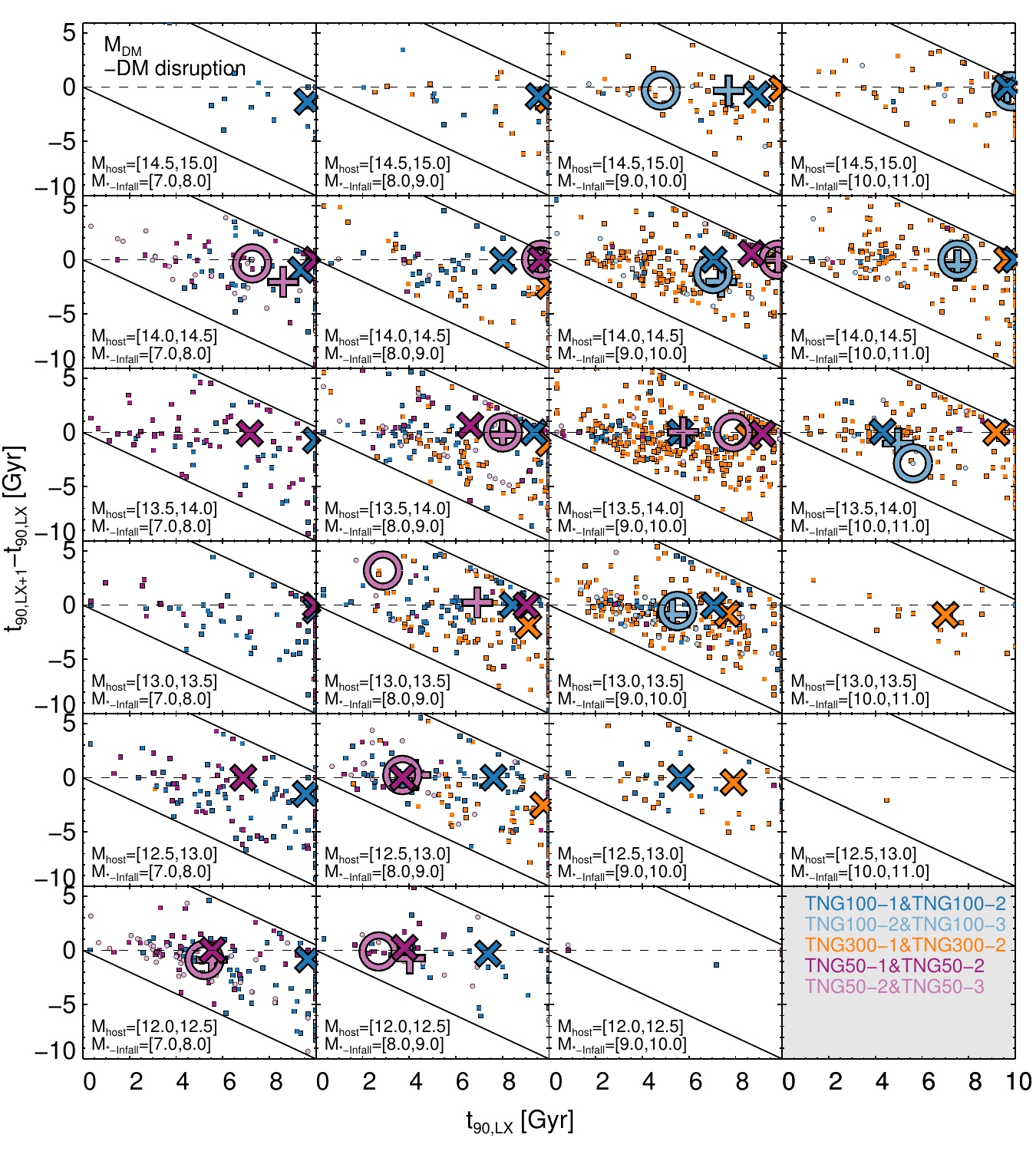}
   \caption{Same as Fig.~\ref{fig:STDM1} but for the time after infall at which galaxies are first stripped to 10~per~cent (instead of 50~per~cent) of their infall DM mass as measured within 30~kpc, $t_\rmn{90}$. Annotations are as in Fig.~\ref{fig:STDM1}: namely, on the $y$-axis we show the difference in stripping time between high- and low-resolution copies of satellites: L3 minus L2 and L2 minus L1. On the other hand, on the $x$-axis we show the $t_\rmn{90}$ of the higher resolution counterpart. Here we focus on the stripping of massively-stripped satellites, to the point where the majority of their mass has been removed. The values of $t_\rmn{90}$ are higher than $t_\rmn{50}$, but are similarly independent of resolution. }
    \label{fig:STDM2}
\end{figure*}

The status of disruption, as opposed to stripping, is a more challenging target for resolution studies, given that in its most effective definition it considers the point at which the satellite is no longer detected in the simulation. We therefore opt to measure the time at which the DM mass has been stripped to 10~per~cent of its original value, $t_\rmn{90}$, at least as a proxy for the mass at which the distribution of mass in the satellite has been strongly disturbed. We adopt the same procedure as Fig.~\ref{fig:STDM1}. We expect that stochastic alterations to orbits of satellites after pericentre for different resolution versions would reduce the value of one-to-one matches. We present the results in Fig.~\ref{fig:STDM2}. 

The first result to note is that many satellites survive with more than 10~per~cent of their infall mass right through until the end of the simulation at $z=0$, as demonstrated by instances where the median symbols are located beyond 9~Gyr. The exceptions to this case are the bins of high satellite mass and low host mass (high $M_\rmn{*-Infall}$--low $M_\rmn{host}$, bottom right panels), where the merger timescales are shortest and therefore a majority of L1 galaxies are disrupted before 8~Gyr. The evidence for resolution dependence in ``disruption'' is stronger than for stripping, although we note that, of the two, disruption is more difficult to define unambiguously. The TNG300-2 satellites are on average disrupted, i.e. lose 90 per cent of their initial DM mass, 1-2~Gyr before their TNG300-1 counterparts (orange crosses). A similar delay occurs for TNG100-3 and TNG100-2 (turquoise plus signs and circles) satellites, which crucially have the same mass resolutions as TNG300-2 and TNG300-1 respectively. A similar, strong dependence of severe stripping on resolution was obtained by \citet{Martin24}. However, the L2-$M_\rmn{*-Infall}$ binned TNG100-2 versus TNG100-3 data show disruption timescales more similar to their L1-$M_\rmn{*-Infall}$ binned equivalent than to TNG300-2 versus TNG300-2, which implies that the difference between TNG300-1 and TNG100-2 is again affected by differences in the available halo sample, as we found for $t_{50}$.

For higher resolution pairs with $M_\rmn{*-Infall}>10^{9}$~$\msun$ -- namely, TNG100-2, TNG100-1 and better -- the difference reduces to zero at the per~cent level in the aggregate. The L2-$M_\rmn{*-Infall}$ binned TNG100-2 versus TNG100-3 data track the L1-$M_\rmn{*-Infall}$ binned version rather than TNG100-1--TNG100-2, which again implies that stochastic halo assembly histories are relevant. Therefore, the disruption times of satellites can be considered to have converged in the aggregate for simulations of the resolution TNG100-2 / TNG300-1 ($\gsim1000$ DM particles) in the mass ranges $M_\rmn{*-Infall}$--$M_\rmn{host}$ range considered.

\subsubsection{Stripping times of stellar mass}
\label{sec:pst_stars}

\begin{figure*}
    \centering
     \includegraphics[scale=0.7]{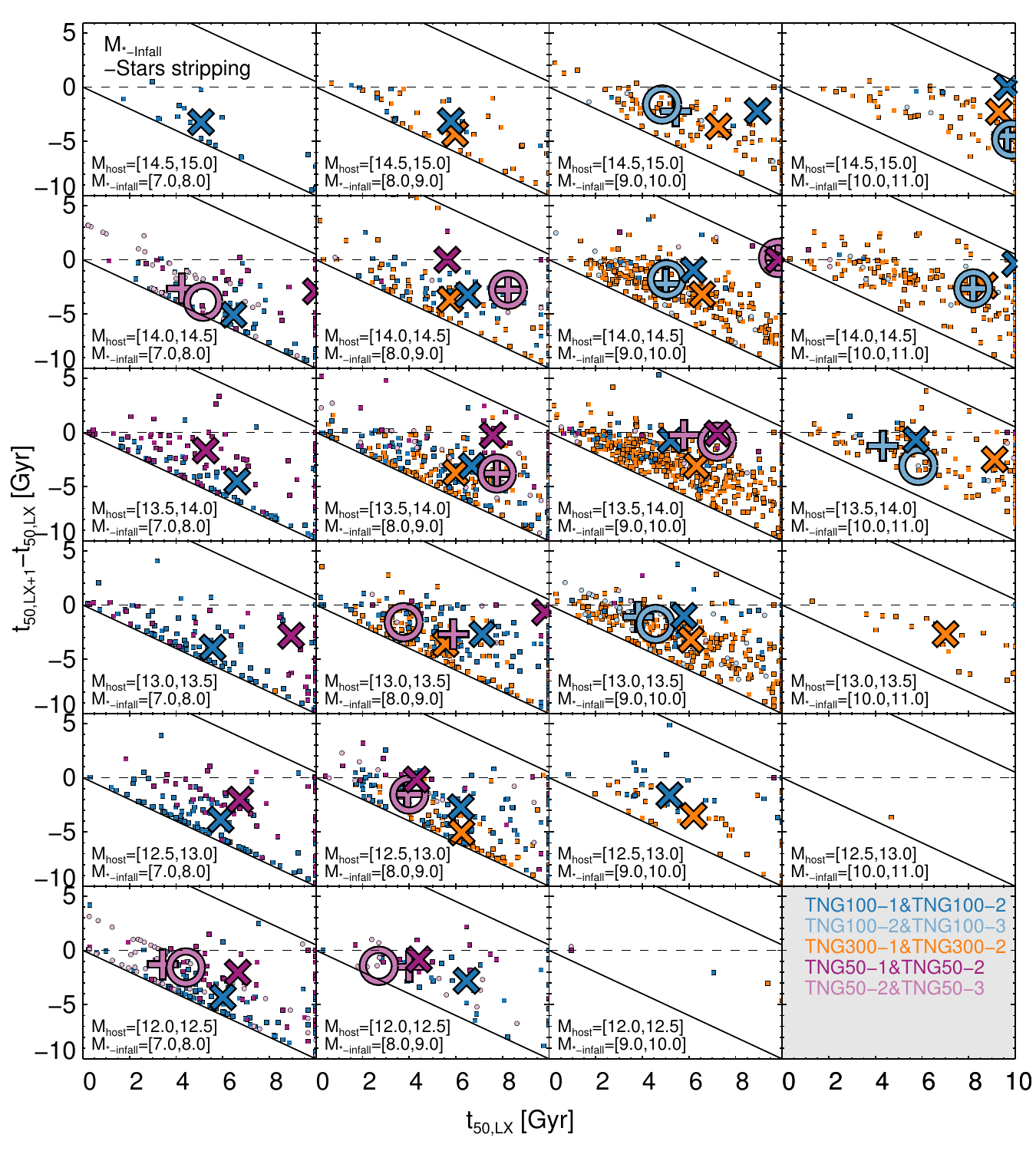}
      \caption{This figure repeats the analysis of  Figs.~\ref{fig:STDM1} and \ref{fig:STDM2} for the stripping of stellar mass. We show the time after infall at which galaxies are first stripped to 50~per~cent of their infall stellar mass, as measured within 30~kpc, $t_\rmn{50}$. We only use star particles that were formed prior to infall (see Table~\ref{tab:def}). Annotations and definitions are the same as in Figs.~\ref{fig:STDM1} and \ref{fig:STDM2}. Both in the averages and for the individual satellites, the rate at which stellar mass is stripped is more prone to resolution effects than that of DM, especially if the resolution is significantly worse than a baryonic mass resolution of $10^{6-7}\msun$ (i.e. worse than TNG100-1 and TNG300-1).}
          \label{fig:STiStar1}
\end{figure*}

\begin{figure*}
    \includegraphics[scale=0.7]{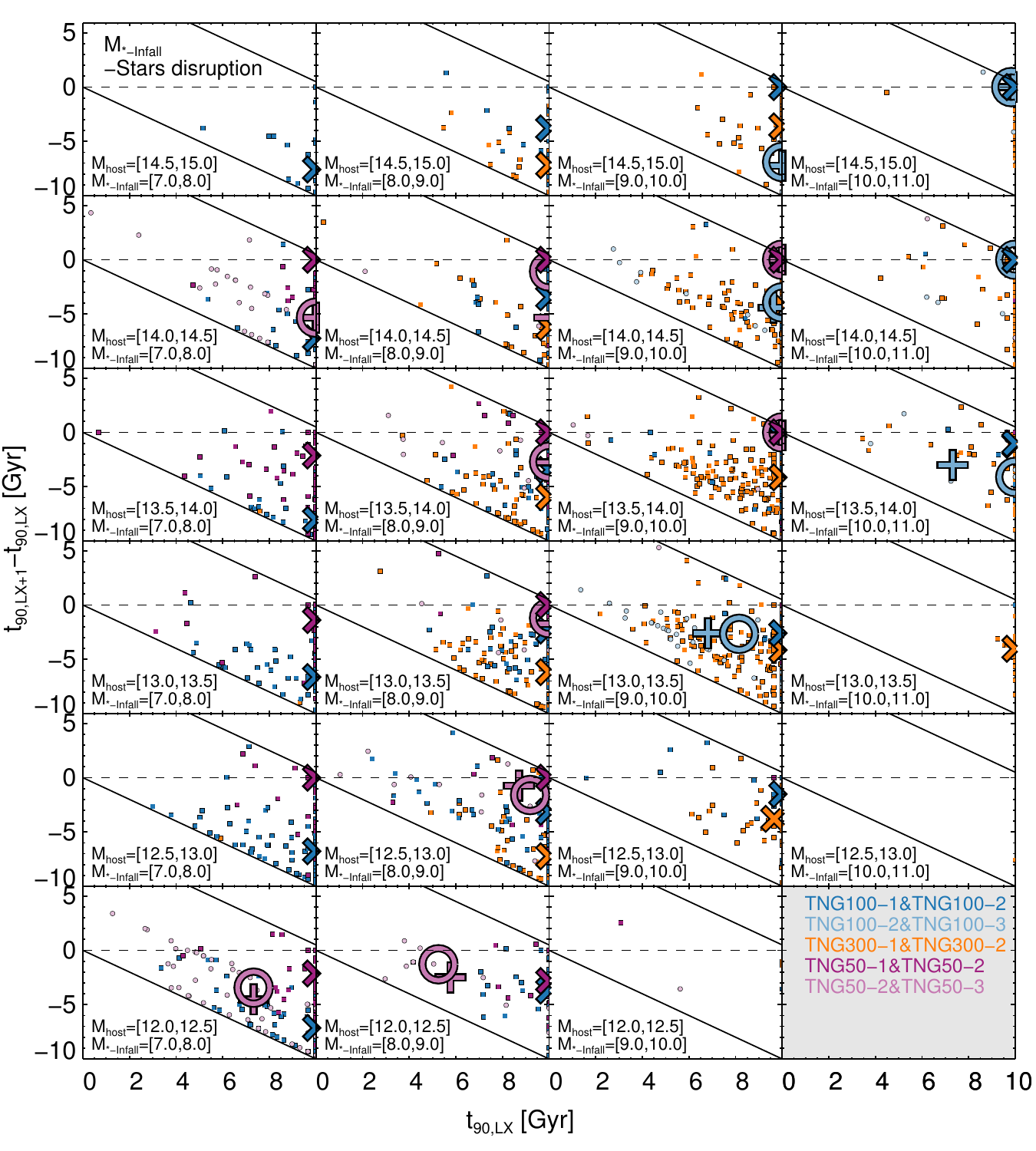}
    \caption{As Fig.~\ref{fig:STiStar1} but focusing on when satellites become massively stripped. Namely, we show the time after infall at which galaxies are first stripped to 10~per~cent (instead of to 50 per cent) of their infall stellar mass, as measured within 30~kpc, $t_\rmn{90}$. Annotations and colours are as Fig.~\ref{fig:STiStar1} and again we only consider star particles that formed prior to infall. Note that a majority of higher-resolution galaxies are not stripped below 10~per~cent of their infall stellar mass across the considered time span of 10 billion years, therefore the symbols are located at the very edge of the plotting area.}
    \label{fig:STiStar2}
\end{figure*}

We have demonstrated the degree to which resolution impacts the DM mass of the halo and here transition to the effect on the stellar component, with a view to understanding both the distribution of satellites and the times at which their stars may be deposited in the hosts' stellar halo. We repeat the exercise from Fig.~\ref{fig:STDM1} to obtain differences in the stellar mass stripping time to 50~per~cent of the infall value -- still excluding any star particles formed after infall --  and present the results in Fig.~\ref{fig:STiStar1}. 

The distribution of stripping times in stellar mass is shifted relative to those of DM mass, from typically 1-2~Gyr after infall for DM to $\sim5$~Gyr for the stars. This result reflects the greater concentration of the stellar component than the DM component. These stellar mass stripping times are similar to the DM disruption times, i.e. the time required to strip to 10~per~cent of the infall DM mass is the same as to strip 50~per~cent of the stellar mass. The dependence on resolution is very evident, with TNG300-2 galaxies stripped on average 4~Gyr earlier than TNG300-1 counterparts across all bins (orange crosses), and a time difference of 3~Gyr between TNG100-2 and TNG100-1 (blue crosses). Only TNG50-2 versus TNG50-1 consistently shows zero time difference at the $<10$~per~cent (purple crosses), with the exception of the $M_\rmn{*-Infall}=[10^{7},10^{8}]$~$\msun$ bins where the L2 counterparts are stripped on average 2-3~Gyr before L1. The significant difference between L2--$M_\rmn{*-Infall}$ binned satellites and their counterpart resolutions in higher volumes implies that stochastic effects remain very important for stellar mass stripping. We therefore conclude that achieving convergence in the stripping of stellar mass is much more challenging than is the case for DM: one requires at least 1000 DM particles in the satellite for $M_\rmn{*-Infall}\sim10^{8}$~$\msun$ hosts and, somewhat paradoxically, even more DM particles for the most massive satellites. This result may be due, at least in part, to the lower concentration of both the DM and stellar components at worse resolution, which we discuss in Section~\ref{sub:conc}.

Finally, we consider the disruption time of the stellar components, defined once again as the time to strip the stellar mass to 10~per~cent of its infall value. Given the greater concentration of the stars compared to the DM, this version of $t_\rmn{90}$ is a particularly interesting measure of when a satellite has been completely disrupted. We construct $t_\rmn{90}$  distributions in the same manner as for Fig.~\ref{fig:STDM2} and show these results in Fig.~\ref{fig:STiStar2}.

A clear majority of most of the satellites' stellar components survive through the full 10~Gyr, indicating that most satellites are not in fact fully disrupted. TNG300-2/TNG300-1 (orange crosses) and TNG100-3/TNG100-2 (turquoise plus signs) galaxies do show a consistent median difference in disruption time \citep[see also][]{Alonso23}, which is longer for less massive satellites; however, importantly, there is a strong trend towards convergence such that TNG100-2 is converged at the per~cent level with respect to TNG100-1 at $M_\rmn{*-Infall}>10^{9}$~$\msun$ (blue crosses) and TNG50-2 at TNG50-1 at $M_\rmn{*-Infall}>10^{8}$~$\msun$ (purple crosses). Given that TNG50-2 has the same resolution as TNG100-1, the TNG100-1 simulation is likewise converged. We note that more than half of the galaxies in our TNG100-1 galaxy sample survive to $z=0$, and therefore we mostly do not have true disruption in these environments, even if a satellite loses more than 90~per~cent of its DM mass within 30~kpc. In summary, the disruption rate is approximately converged for resolutions of TNG100-1 and better.  

Having shown that disruption time is approximately independent of resolution, we now discuss whether this disruption is physically realistic or instead represents convergence to an unphysical timescale as argued by \citet{vdBosch18}. Their paper argues that satellites on idealised circular orbits should be stripped towards some asymptotic value over a 10~Gyr period. Instead, the combination of spatial resolution and mass resolution employed in cosmological simulations leads to spurious disruption well before 10~Gyr, and that the choice of scaling spatial and mass resolution leads to the same timing of disruption.

In their model, satellites on circular orbits lose 90~per~cent of their mass within 2~Gyr of infall and more than 99~per~cent of their mass within 10~Gyr, such that they would be disrupted below the resolution limit in the `fiducial' softening resolution; by contrast, a majority of our satellites survive for more than 10~Gyr when using the stellar 10~per~cent stripping time as a proxy for the satellite halo disruption time. We expect that the crucial difference is the eccentricity of orbits in cosmological simulations compared to the circular orbits of \citet{vdBosch18}. In fact, \citet{Errani20} showed explicitly with idealised simulations that for increasing eccentricity the stripping rate decreases \citep[see also discussions in][]{Bahe19}. We compute (although do not show) the ratio of first apocentre distance to first pericentre distance for our TNG satellites, and find that 84~per~cent of the satellites exhibit an apocentre-to-pericentre distance ratio of 2. In the approximation of elliptical orbits, the corresponding eccentricity of this 84~per~cent of the data is 0.67 and is therefore very different to the circular orbits of \citet{vdBosch18} \footnote{In addition to being circular, the fiducial orbits used by \citet{vdBosch18} have a radius 50~per~cent of the host halo virial radius. A careful comparison to the \citet{vdBosch18} result would also require taking into account the orbit radius as well as the ellipticity; we leave this analysis to future work.}.  

 \begin{figure}
     \centering
     \includegraphics[scale=0.5]{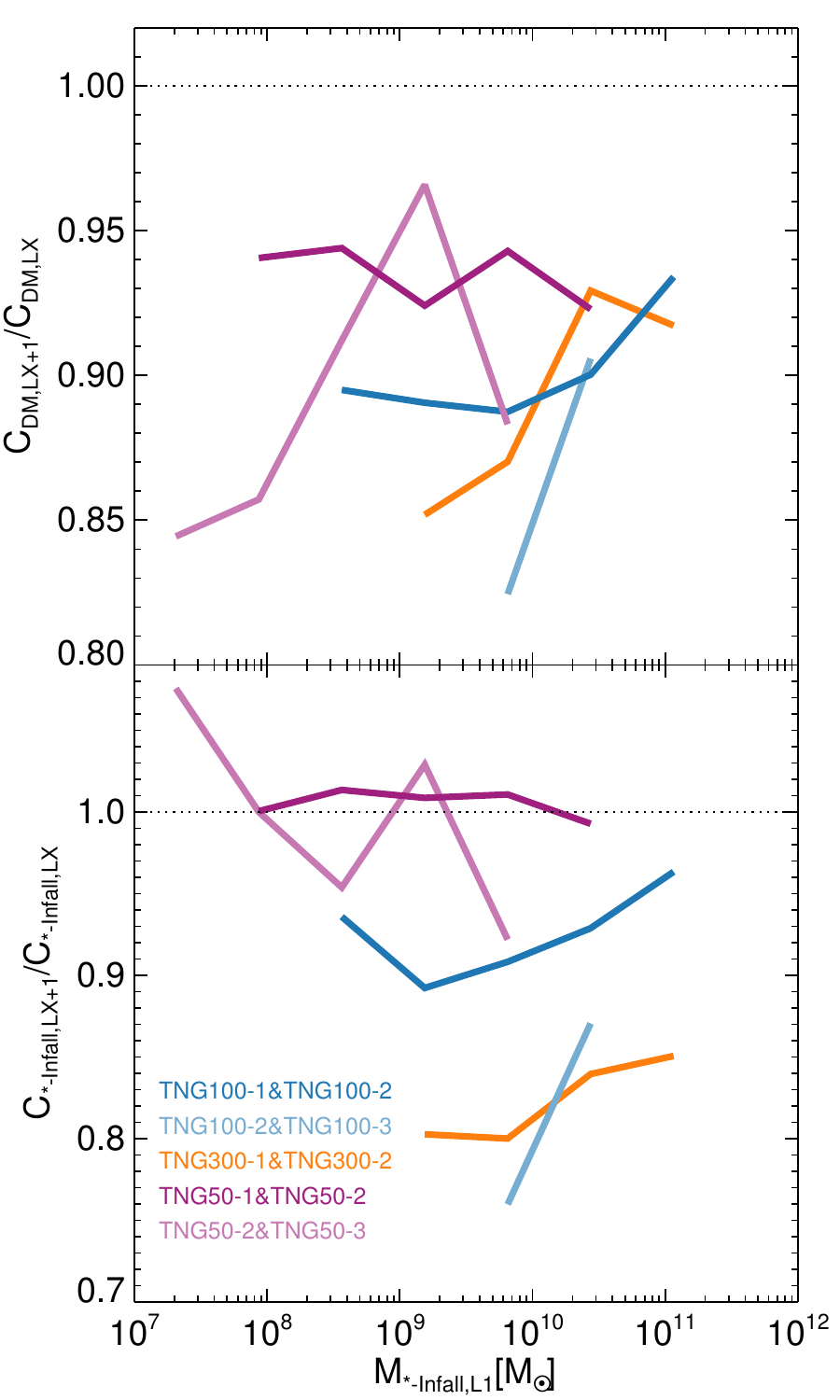}
     \caption{The median ratio between resolution levels (L2/L1 and L3/L2) of the mass concentrations of the galaxies at infall, as a function of infall stellar mass at the best (L1) resolution level. In this case concentration refers to the ratio of the mass within 5 kpc to that within 30 kpc (see Table~\ref{tab:def}), for the DM component (top) and the stellar mass (bottom). 
     The correspondence between curve colour and the simulation is given in the figure legend. The 68~per~cent scatter about the median is typically 0.2 points, and is omitted for clarity. We omit the TNG300-3 data due to its very poor resolution as discussed above. Interestingly, the resolution convergence of the compactness of galaxies is better for the stellar than for the DM component (e.g. magenta curves across the two panels). Moreover, at better resolution, the concentrations of galaxies are approaching convergence and they are typically captured at the resolution of TNG100-1 to better than 10-15~per~cent.}
     \label{fig:CP}
 \end{figure}
 
Were spurious disruption to take place for our satellites, even at these much later times, we would expect that the difference in disruption would be zero for all sufficiently resolved satellites, and this is not the case. Even though the disruption times are resolved in the {\it aggregate} -- as represented by the median -- on a satellite-by-satellite case the scatter is very large indeed, which is likely due to the stochastic changes to orbits as shown in Fig.~\ref{fig:IEDM}. We therefore conclude that the statistics of the general satellite populations are not influenced strongly by spurious disruption: our distribution of orbits is more robust to stripping, and stochastic scatter over many Gyr will wash out any imprint of spurious disruption. The question remains of whether disruption has an effect on the specific set of satellites characterised by very radial orbits \citep{Grand21}; we defer this analysis to future work.

In conclusion, we have shown that the stripping of the satellites' DM component does not vary systematically with resolution except where the resolution is particularly poor. Namely, satellite disruption correlates with $m_\rmn{baryon}$ for $m_\rmn{baryon}>10^{8}$~$\msun$, (e.g. TNG300-3, which was indeed designed for resolution studies and not for science) but we do not find evidence that the disruption rates are affected by spurious disruption of the kind discussed in \citet{vdBosch18}. After first pericentre the orbits scatter and therefore the relationship between disruption time and numerical resolution becomes less clear. By contrast, the evolution of the stellar mass stripping rates is strongly resolution dependent, although the inner core of stars is sufficiently dense to survive through the full 10~Gyr time period at most resolution levels.

\subsection{The distribution of matter within satellites}
 \label{sub:conc}
 
 The stripping rate of stars from a satellite galaxy depends on the distribution of both its DM and its stars at infall. The DM is the dominant matter component, and denser DM distributions will make their stellar populations more resistant to stripping. The same is true to a degree for the stellar mass. If the distribution of stars is very extended, it follows that as the satellite tidal radius shrinks with decreasing distance to the halo centre, the tidal radius will quickly become smaller than the stellar radius, and thus more diffuse satellites will experience more stripping at greater distances from the host centre.
 
The description of the matter distribution needs to reflect the relative abundance of material in the satellite's central region as compared to the satellite as a whole. We define the central region as the material $<5$~kpc from the satellite centre of potential, and the galaxy as a whole as 30~kpc from the same centre of potential. We hence compute the concentration, $C_{X}$, where $X$ is the matter species, as the ratio of the $X$ mass within 5~kpc of the satellite centre to the $X$ mass within 30~kpc. We compute this quantity for all of our satellites at infall, and then calculate the ratio of $C_{X}$ from low resolution to high resolution satellites: if this ratio is lower than 1 then the high resolution satellite is more concentrated than its low resolution counterpart. We compute the median ratio in low resolution-high resolution satellites as a function of the L1 infall stellar mass, and present the results in Fig.~\ref{fig:CP}.

The DM concentrations are more strongly influenced by resolution than the stellar concentrations, as per the top versus bottom panels of Fig.~\ref{fig:CP}. On average, the concentration increases by at least 5~per~cent with each $8\times$ step in mass resolution. We therefore conclude that this quantity is not yet trending towards convergence in our simulations. It is possible that even TNG50-1 is underestimating the satellite central density, although the stellar component makes it unlikely that the rate of stripping would change at higher resolution (see below).  
 
The behaviour of the stellar mass is marginally more complex. There is a clear correlation between resolution and the change in concentration. Between TNG300-3 and TNG300-2 the median change is 40~per~cent, but by TNG50-2 and TNG50-1 there is no measurable change. It is therefore the case that the concentration of both TNG50-2 and TNG100-1 are converged in their stellar concentrations. This shows that the concentrations of both the DM and stellar mass components of pre-infall TNG galaxies correlate with resolution, and therefore higher resolution galaxies may be more resilient to stripping. However, the ways stripping and concentrations change with both numerical resolution and matter species show that the resolution dependence of the $C_\mathrm{*,infall}$ does not fully justify the resolution dependence of stellar mass stripping. This is because, as we show explicitly below, $M_\mathrm{*,infall}$ itself is not converged.  

\subsection{Ex-situ stellar halo progenitors}
\label{subsec:shpp}

In this subsection we shift to considering the intra-group light and stellar haloes of galaxy groups and $L_{*}$ galaxies, and the satellites from which they are constructed, by focusing exclusively on their accreted, i.e. ex-situ, components.

Changes in resolution will have multiple impacts on how infalling galaxies contribute to the stellar halo. The first factor is the increase in total stellar mass formed for higher resolution, and the second is that the concentration of the stars and DM determines how easily stripping can be resisted. We have already discussed the concentrations above; in this section we instead consider the stellar mass deposited in the host halo. We will therefore demonstrate how the resolution dependence of the satellite stellar mass and concentration affects the stellar halo. We first discuss how we assemble our host galaxy sample and then present our results. 

\subsubsection{Star particle selection}

We begin our analysis by selecting all of the haloes in the three TNG boxes that have masses in one of two mass ranges: $M_\rmn{host}=[10^{13},10^{14}]$~$\msun$ and $M_\rmn{host}=[10^{12},10^{12.5}]$~$\msun$ to compare to the stellar haloes of groups and $L_{*}$ mass galaxies, respectively. For the $L_{*}$ galaxies we select hosts from the TNG100 and TNG50 volumes, whereas for galaxy groups we select hosts in all three volumes. From these hosts we select all star particles that are formed ex-situ \citep{RodriguezGomez16}, in that they formed in another galaxy that was later accreted into the host halo~\footnote{We do not differentiate between ex-situ stars that contribute to a bulge component versus a stellar halo/intra-cluster light component. All accreted stars are included in our initial analysis, and our later analysis of stripping radius will implicitly remove the ex-situ bulge component.}. We subsequently identify the galaxy in which each ex-situ star formed, which we term the `progenitor' galaxy. 

 Each such progenitor galaxy contributes a given amount of stellar mass to the host as identified by {\sc subfind}, and we refer to this mass total as the `stripped stellar mass', $M_\rmn{*-Strip}$. In addition, we use merger trees to identify the peak stellar mass that the progenitor galaxy exhibits over its lifetime: we label this mass $M_\rmn{*-Peak}$. Therefore, if a progenitor galaxy is stripped out of existence by the host and all of its stars are located within the host, $M_\rmn{*-Strip}=M_\rmn{*-Peak}$. Note that the maximum stellar mass can be reached {\it after} infall due to (continued) star formation \citep[e.g.][]{Engler21b}, and that unlike $M_\rmn{*-Infall}$ these definitions use all star particles found to be bound to their halo and not just particles within 30~kpc of the galaxy centre.
 
We use our inter-resolution matching catalogues to identify pairs of progenitor galaxies across resolutions. The catalogue that we use is built for $z=1$ galaxies, therefore we restrict our samples of progenitor galaxies to galaxies that are existent at $z=1$ in both resolution catalogues. It is possible for the progenitors to still be present in the galaxy catalogue at $z=0$; it will have begun the process of transferring mass to the stellar halo of the host at some earlier time. Unlike Section~\ref{subsec:tsg}, we do not include any limits on the infall time of the satellite.     

\subsubsection{Impact of resolution to the build up of stellar haloes}

\begin{figure*}
    \centering
     \includegraphics[scale=0.345]{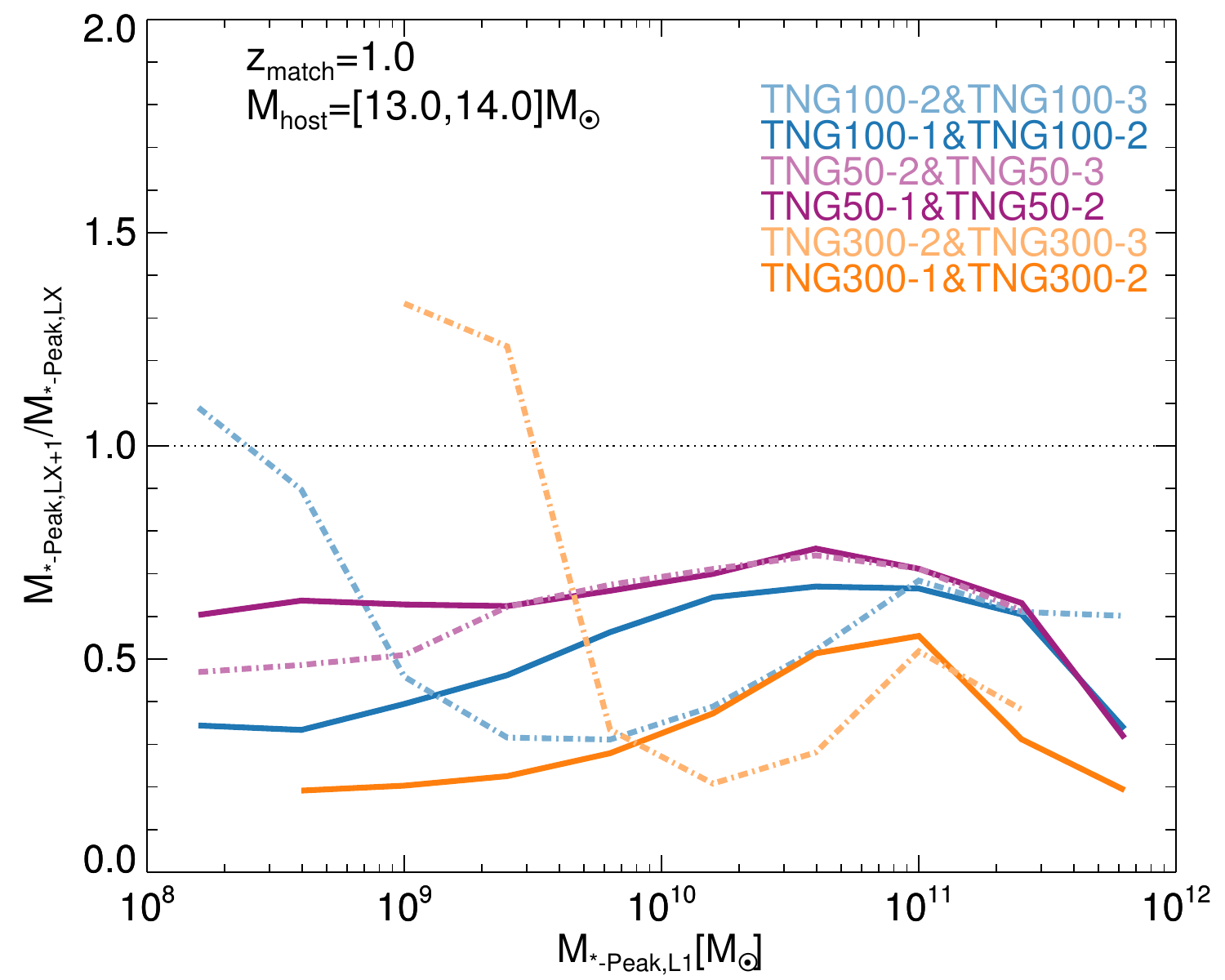}
        \includegraphics[scale=0.345]{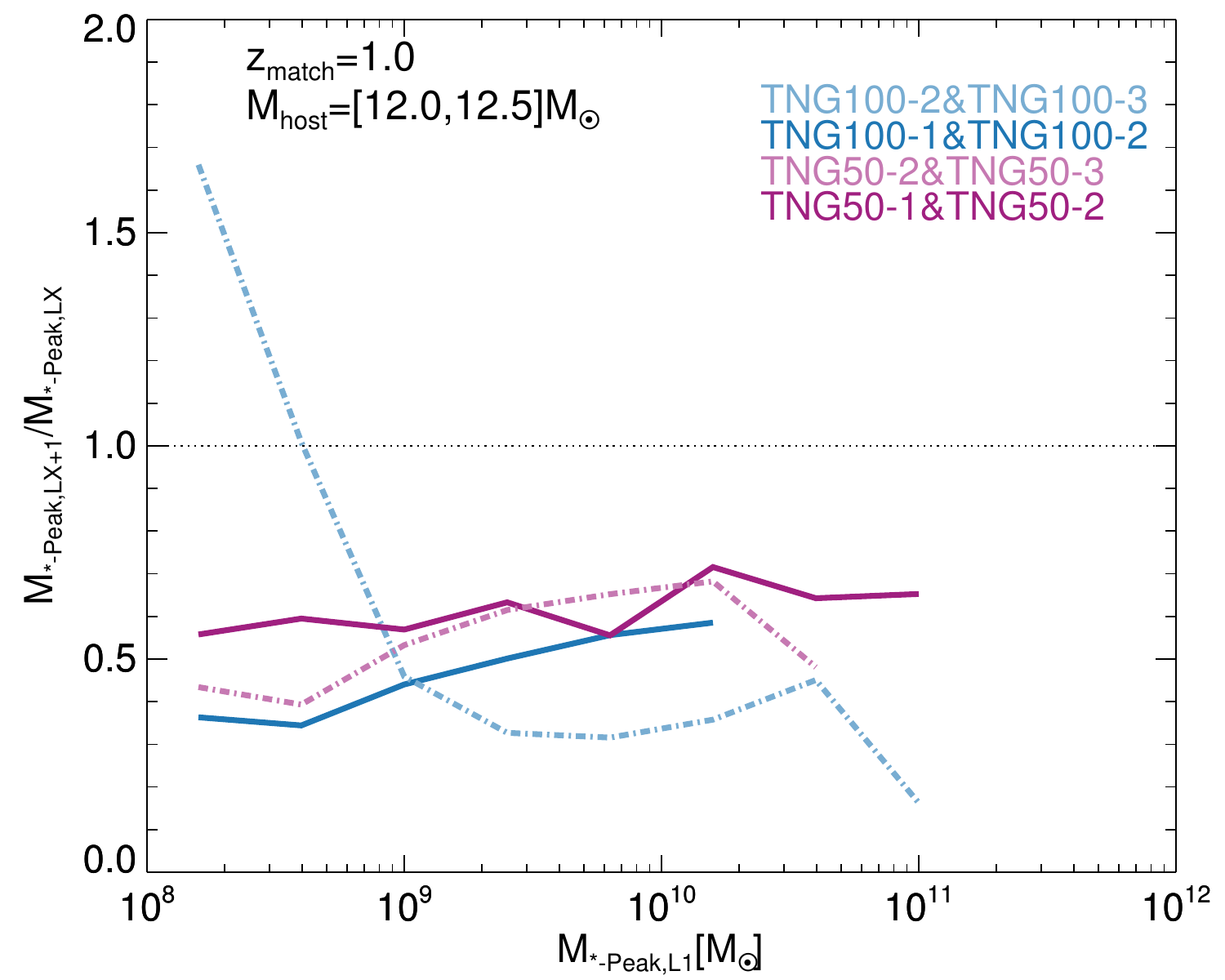}
    \caption{Maximum stellar masses associated with progenitor galaxies that contribute to the stellar haloes in galaxy groups (left-hand panel, for host mass $[10^{13},10^{14}]$~$\msun$) and $L_{*}$ galaxies ( right-hand panel, for $[10^{12.0},10^{12.5}]$) at $z=0$. The median ratio of each galaxy's low resolution to high resolution (L3--to--L2 and L2--to--L1) maximum stellar mass, $M_\rmn{*-Peak}$, is plotted as a function of the L1 resolution counterpart. The line colours denote the counterpart simulations, as given in the figure legend. All galaxies are present at $z=1$ by construction, and may or may not survive until the present day. TNG300 data are only shown in the left-hand panel; TNG100 and TNG50 results appear in both panels. Convergence improves for better resolution but not to no better than 30~per~cent between TNG50-1 and TNG50-2.}
    \label{fig:HSProgs}
\end{figure*}

We first consider the effect of the change in resolution on the progenitor maximum mass. We compute the ratio of the maximum stellar mass $M_\rmn{*-Peak}$ between the low and high resolution counterparts, and then calculate the median ratio as a function of the maximum mass in the L1 counterpart. We perform this algorithm for L3/L2 pairs and L2/L1 pairs. In both cases we use the L1 counterpart to mark the position on the $x$-axis so that for the data in each panel changes in resolution are only reflected in the change in $y$-axis position. We present the results for both halo mass bins in Fig.~\ref{fig:HSProgs}.

We recover the result from previous TNG papers and from Section~\ref{sec:subabund} that the stellar mass increases as the simulation mass resolution improves \citep{Pillepich17b,Engler21}. In general, the suppression in $M_\rmn{*-Peak}$ is stronger towards lower masses. Taking TNG100 as an example, the $M_\rmn{*-Peak}\sim3\times10^{10}$~$\msun$ stellar masses are suppressed by 45~per~cent at L2 relative to L1 and 70~per~cent at L3 relative to L2. The scaling is somewhat more consistent for the same galaxy mass in TNG50, where the step in stellar mass is 30~per~cent from L1 to L2 and a further 30~per~cent from L2 to L3. The sharp up-turns at low mass are regions where the low number of particles causes sampling issues. We find approximately the same results for the two halo mass bins except for the absence of massive objects in the $L_{*}$ galaxy haloes.

\begin{figure*}
    \centering
     \includegraphics[scale=0.435]{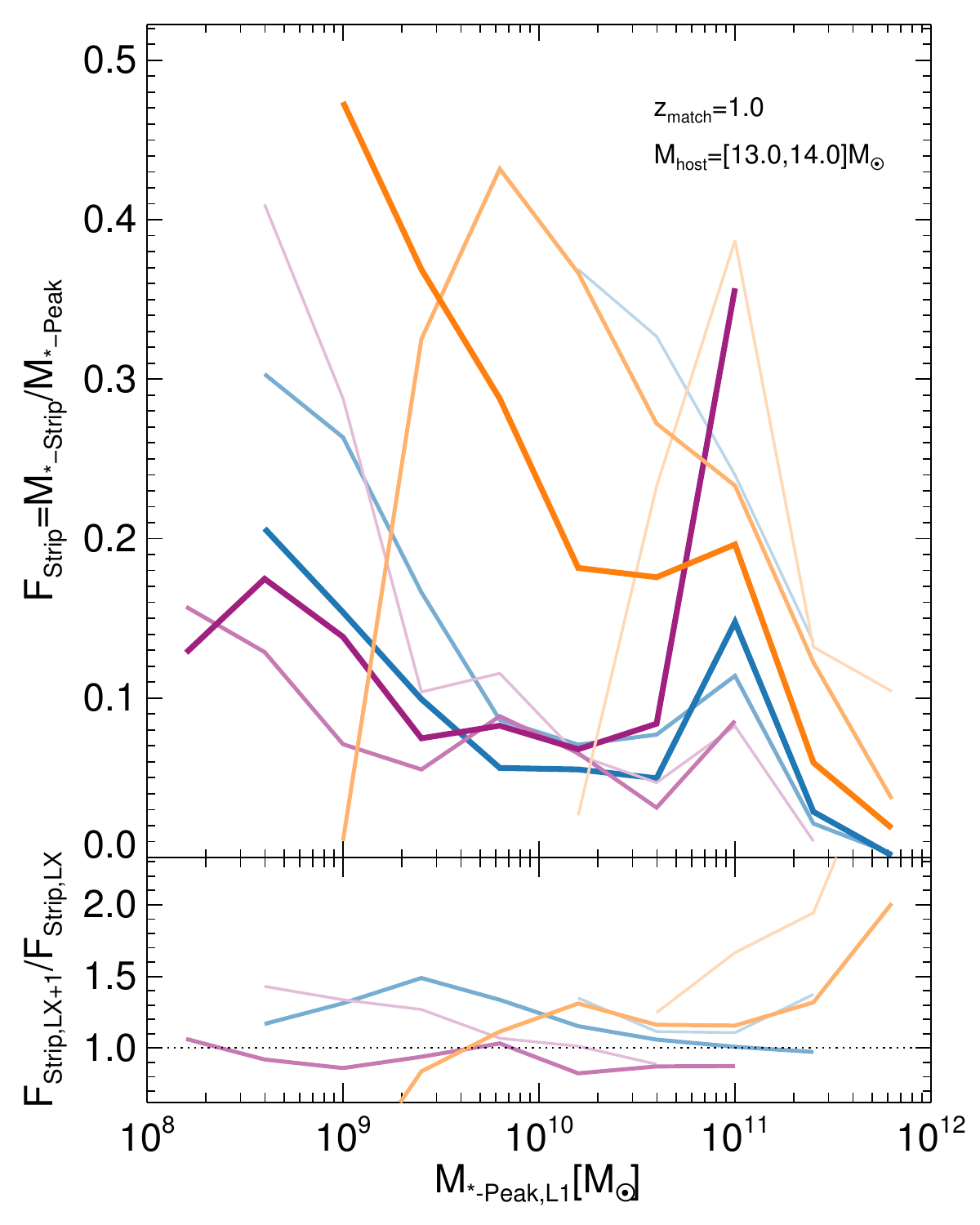}
       \includegraphics[scale=0.435]{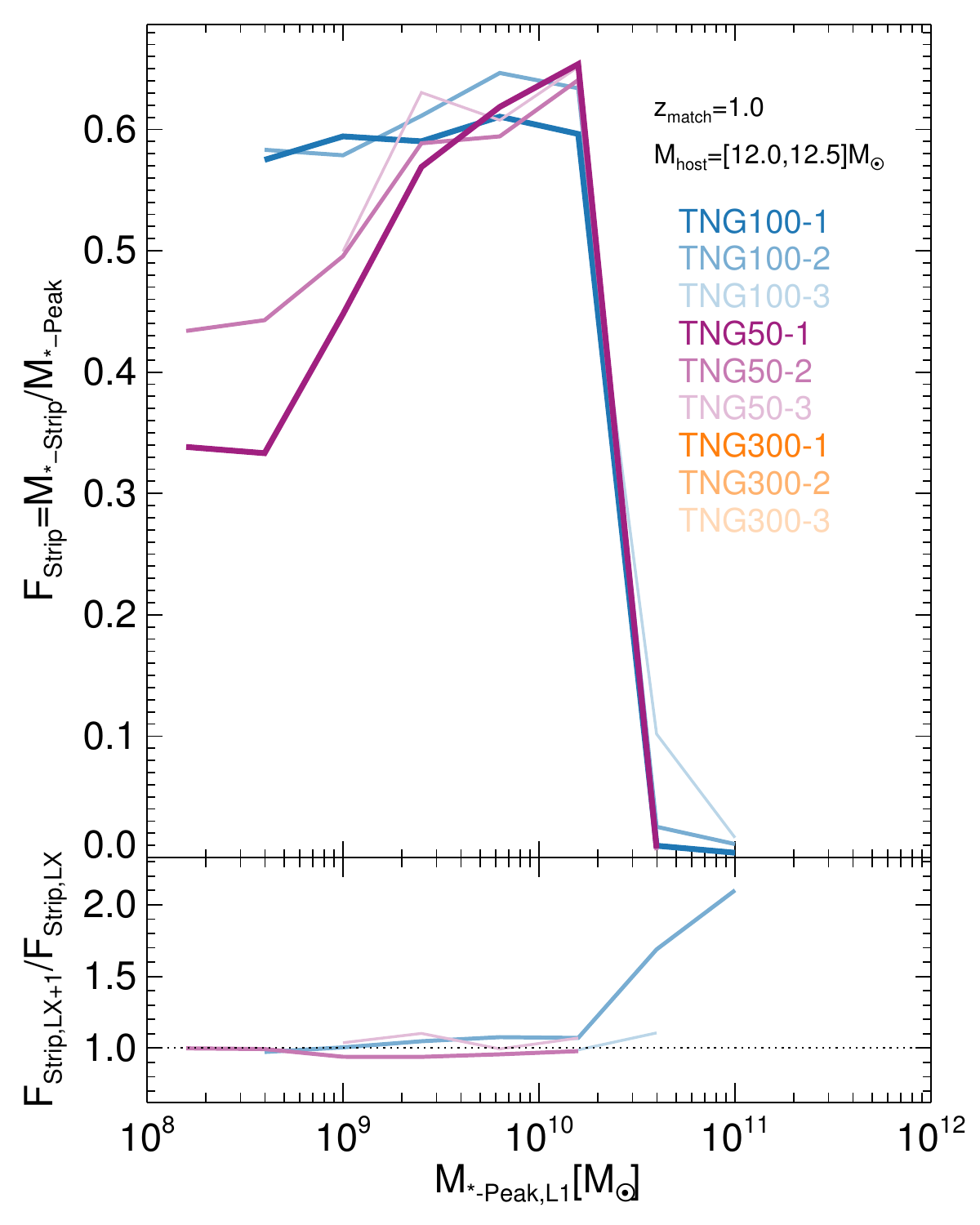}
    \caption{The satellite stellar mass stripping fraction around group-mass hosts (left) and L* galaxies (right), which will go on to build the ex-situ stellar halo. Namely, we show the median fraction of stellar mass that is stripped from the progenitor maximum mass into the host stellar halo. The colour scheme is the same as in Fig.~\ref{fig:HSProgs} but now the L2 and L3 data are plotted with thinner lines to improve legibility. 
    In particular, in the top panels, we plot the median mass fractions for each simulation as a function of the L1 counterpart maximum mass, and in the bottom panels we plot the median ratio of the fractions between adjacent resolution levels. Note that this quantity is not equivalent to the ratio of the curves displayed in the top panel. The degree of convergence in these stripping fractions is much better than for stellar mass, especiallly for $L_{*}$ hosts.}
    \label{fig:HSProgsRat}
\end{figure*}

We can interpret the likely impact on the stellar haloes of host galaxies by studying the stripping rates between counterpart galaxies. We consider the ratio of $M_\rmn{*-Strip}$ and $M_\rmn{*-Peak}$, $F_\rmn{s}$, which is the fraction of each progenitor galaxy's stellar mass that is donated to the host stellar halo; note that both mass definitions include any stars that were formed in post-infall starbursts. We compute $F_\rmn{s}=M_\rmn{*-Strip}/M_\rmn{*-Peak}$ for each of our progenitor galaxies and present the median ratios as a function of the L1 counterpart $M_\rmn{*-Peak}$ in Fig.~\ref{fig:HSProgsRat}, with one panel for the group mass haloes and a second for the $L_{*}$ hosts. We include all L3 galaxies and L2 galaxies that have a counterpart in L1, and the L1 galaxies that have a counterpart in L2.

We begin our analysis with the group-mass haloes. The strongest stripping rates are observed in two classes of objects: massive satellites ($M_\rmn{*-Peak,1}>3\times10^{10}$~$\msun$) and satellites in simulations of DM particle mass $\geq5.9\times10^{7}$~$\msun$ (TNG100-2, TNG300-1 and lower resolution). The strong stripping of bright galaxies can be explained by the short merger times of massive galaxies under dynamical friction; given that these objects are rare, the degree of stripping does not correlate with resolution. For less massive, low resolution galaxies, the stripped proportion can be a factor of two higher than at high resolution: at $M_\rmn{*-Peak,1}=10^{10}$~$\msun$ the TNG300-2 galaxy population experiences twice as much stripping as TNG300-1, and TNG100-2 some 30~per~cent more stripping than TNG100-1. We can explain this result from the correlation between resolution and concentrations: both the stellar and DM components increase in density with resolution and are therefore (i) more resistant to stripping and (ii) have a greater fraction of their stellar mass within the tidal radius. 

We further explore this argument by computing the ratio of galaxy pairs and plotting the median ratio of the stripping fraction as a function of L1 maximum mass. The median stripping fraction ratio in satellites $M_\rmn{*-Peak,1}>10^{11}$~$\msun$ is greater than 1.2 for simulations of the resolution of TNG100-3 and worse. There is some evidence for satellites of mass $M_\rmn{*-Peak,1}\sim2\times10^{9}$~$\msun$ to show an increase in stripping rates for high resolution of up to 50~per~cent. We speculate that this is a feature of resolution dependence in the stripping rate on either side of $M_\rmn{*-Peak,1}\sim5\times10^{10}$~$\msun$, and is related to the two different concentration relations: that the lower concentration in stellar mass increases the stripping rate in large, high stellar mass galaxies whereas poor DM resolution plays a more important role in low mass galaxies. However, we note that at the best available resolution pair -- TNG50-1 and TNG50-2 -- it is the high resolution simulation that donates on average 10~per~cent more of its stellar mass to the host, across the whole range in $M_\rmn{*-Peak,1}$. This could be a result of a statistical fluke, especially when the stellar concentration on TNG50-1 and TNG50-2 are nearly identical, and given also that this finding is not replicated in the other host mass bin (see below).

We repeat this process for satellites that are stripped by the $L_{*}$ galaxy hosts. The stripping rates are much stronger than in the group mass haloes, at 30 to 60~per~cent, which is due to the shorter dynamical friction times associated with less massive hosts. The peak in stripping occurs at $M_\rmn{*-Peak,1}=10^{10}$~$\msun$; we expect that the small number of galaxies more massive than this limit are only recently accreted, thus explaining their negligible stripping. The ratio in stripping rates between galaxy pairs is close to unity, which suggests that the results have approximately converged. Taking the evidence of both group and $L_{*}$ haloes together, we conclude that TNG50-2 simulation stripping rate has converged with TNG50-1 to better than 10~per~cent for satellites of infall stellar mass $>10^{8}$~$\msun$. Given that TNG50-2 has the same resolution as TNG100-1, we find that the TNG100-1 results are stable to resolution so far as satellite stripping is concerned, and different outcomes at this resolution or better can only be produced by changing the galaxy formation model.

Finally, we consider the impact of resolution on the distribution of stripped stars within the host halo. We have shown that stripping occurs of order 1 Gyr sooner after infall at low resolution than at high resolution. In order to explain the discrepancy measured by \citet{Merritt20}  -- that the TNG model overpredicts the amount of stellar mass at radii $\gsim0.1$~$\RTWOC$ -- with resolution alone, we would have to show that any {\it increase} in the concentration of the stellar halo demonstrated in Fig.~\ref{fig:sh} -- i.e. a steepening of the slope of the stellar halo profile -- is accompanied by a {\it decrease} in the total amount of stars at radii $\gsim0.1$~$\RTWOC$ -- i.e. a lower amplitude of the halo profile at large radii. This hypothesis would be satisfied immediately if the total stellar mass were the same between resolutions; however, we showed in Fig.~\ref{fig:satmf} and Fig.~\ref{fig:HSProgs} that stellar mass increases with improved resolution, so it is not inevitable that a higher concentration will go hand in hand with a fainter outer stellar halo.   

\begin{figure*}
    \centering
    \includegraphics[scale=0.6]{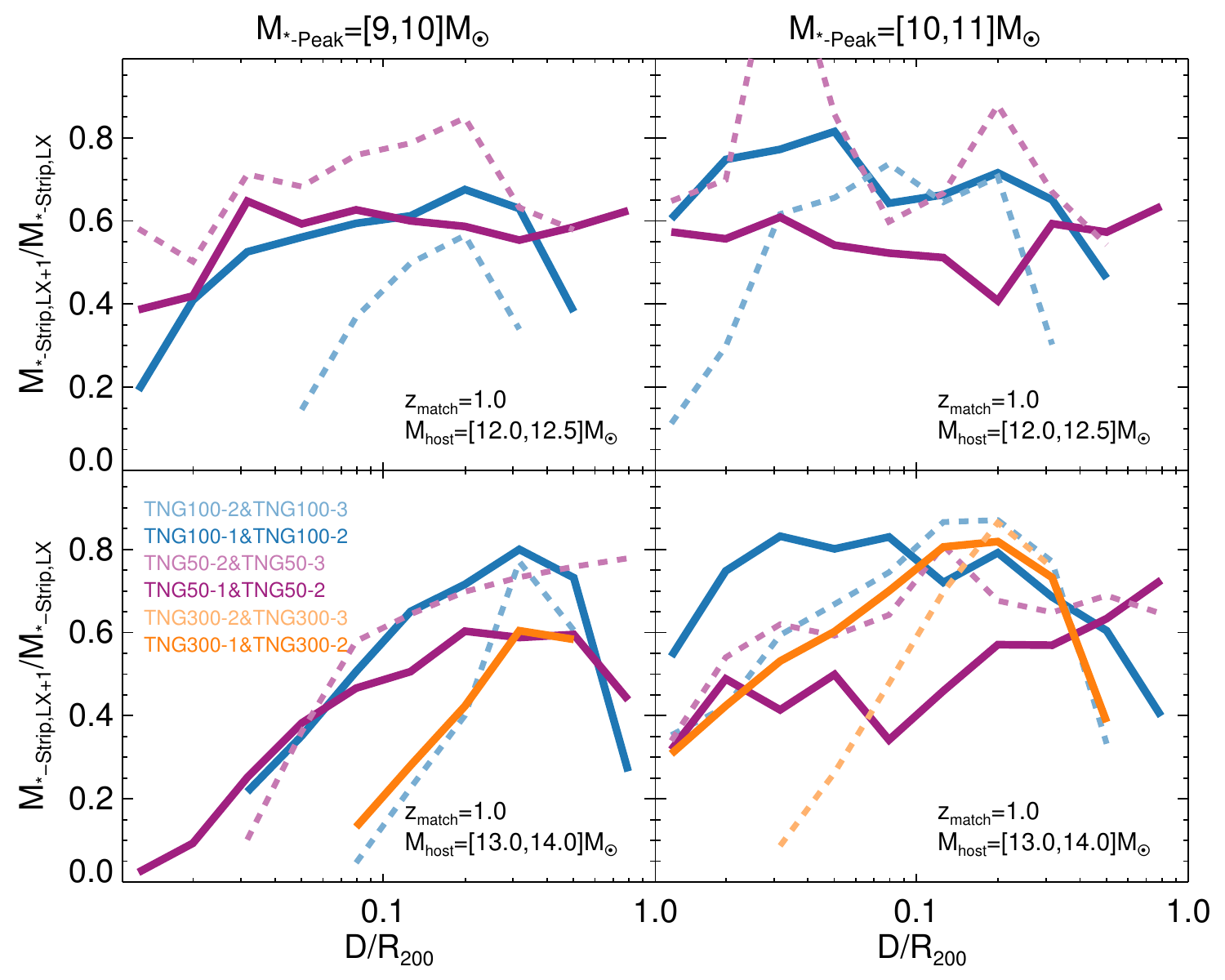}
    \caption{The median ratio of the stellar mass deposited by resolution counterparts as a function of the radius at which it was stripped. We focus on L* hosts in the top panels and on group-mass hosts in the bottom pnales, with different selections of galaxies that bring accreted mass between left and right based on the maximum stellar mass. The relationship between colour and data set is shown in the figure legend. Note that the bottom panels, for L2-L3 pairs we use the L2 $M_\rmn{*-Peak}$ and not the L1 on the x axis. Consistent with the time-based results of the previous figures accreted stars are deposited relatively closer to the hosts' centres as resolution improves. However, the trends with resolution jumps are not strictly monotonic, which we ascribe to stochasticity and low-number statistics.}
    
    \label{fig:HSDist}
\end{figure*}

A careful analysis to test these two hypotheses would involve tracking the locations of individual star particles from the moment of stripping to the present day, which is beyond the scope of this study. We instead take the following approach. We identify the distance from the host halo at which each star particle was registered as stripped from its progenitor, and compute a histogram of stellar stripping distances for each progenitor in units of the host present day $\RTWOC$. We compute the ratio of histograms between low resolution and high resolution counterparts of the progenitor -- L1-L2 and L2-L3 -- and compute the median ratio in each radius bin across progenitors. We present our results for progenitors in group-mass and MW-mass hosts in Fig.~\ref{fig:HSDist}. For both sets of hosts we use two panels: one for $M_\rmn{*-Peak}=[10^{9},10^{10}]$~$\msun$ and a second for $M_\rmn{*-Peak}=[10^{10},10^{11}]$~$\msun$. The value of $M_\rmn{*-Peak}$ is that of the higher resolution progenitor in each pair: L1 for L1-L2 and L2 for L2-L3. 

There is a common feature across almost all panels and data sets for the median stellar mass deposition ratio to decrease at smaller radii. Improvements to the resolution bring the ratios slightly closer to unity and also to extend closer into the halo. There is a correspondence between different volumes that use the same resolution (TNG100-1--TNG100-2 versus TNG50-2--TNG50-3, plus TNG300-1--TNG300-2 versus TNG100-2--TNG100-3) at the 10~per~cent level. The resolution dependence of the mean stripping distance roughly traces a combination of the DM and stellar aperture-concentration resolution dependencies shown above. It is therefore the case that because better resolution increases stellar mass at the same time that the stellar concentration is converged (Fig.~\ref{fig:CP}) the stripping rate is conserved, but the total stellar halo mass is not converged. The one case where stellar mass deposition ratios are not strongly correlated with radius is the set of TNG100 and TNG50 galaxies in the highest mass bin, where it is possible that the combination of the large number particles per progenitor halo and the shorter dynamical friction times wash out some of the resolution effects. We propose that this may be part of the reason for the good agreement between TNG100 and observations found for $\MTWOC=[10^{13.0},10^{13.42}]$~$\msun$ hosts in \citet{Ardila21}.  

We have shown that the stellar halo is indeed more concentrated at better resolutions, provided that the dynamical friction time is long, and therefore confirm the first hypothesis that we made above. However, for our second hypothesis to be satisfied -- that increasing the resolution decreases the stellar mass at large radii -- in most of our progenitors, the median mass ratio would have to be $>1$. Instead, the highest median mass ratio is 0.8: therefore, using better resolution in the TNG model still {\it increases} the stellar mass in the halo outskirts due to the concurrent increase in star formation. 

The most notable difference between the group mass- and galaxy mass-haloes is the profile ratios of $[10^{9},10^{10}]$~$\msun$ haloes, which are flatter for the lower host mass bin, providing further evidence that shorter dynamical friction times -- in satellites massive relative to the host -- wash out any effect of the halo mass profile. However, the equal resolution pairs of TNG100-1--TNG100-2 and TNG50-2--TNG50-3 do not agree with each other in either panel, which implies that scatter between volumes may be affecting our results. We therefore conclude that both host halo classes predict increased stellar mass at large radii with improved mass resolution, and so this cannot resolve the discrepancy between the TNG model and \citet{Merritt20}. 

\section{Summary and conclusions}
    \label{sec:summary}

The assembly of the stellar haloes around galaxies and of the intra-cluster light in groups and clusters, which are mainly made of ex-situ, i.e. accreted, stars, is a strong test of galaxy formation theories. Their proper modeling requires that the following galaxy properties are reasonably well reproduced by any given simulation: the masses and density profiles of satellite galaxies, the stellar density profiles of massive central galaxies, and also the contraction, and more generally properties, of the host galaxy DM halo. In addition to uncertainties concerning which astrophysical processes are relevant for regulating galaxy evolution, model predictions from simulations are necessarily limited by finite numerical resolution, in addition to the unavoidable challenge of a softened gravitational force law. Given that some papers have argued for spurious destruction of subhaloes due to poor resolution \citep{Grand21}, including in ways that cannot be improved by increasing resolution \mbox{\citep{vdBosch18,Green21}}, and that simulations such as TNG are being compared in quantitative detail to low surface-brightness observations \citep{Merritt20, MontenegroTaborda25}, it is necessary to check in what ways numerical resolution may be limiting our ability to study the diffuse light component around galaxies and in groups / clusters of galaxies.

In this paper we used the full range of the IllustrisTNG cosmological simulations of galaxies, namely TNG100-1, TNG300-1, TNG50-1 and their lower resolution counterparts for a total of 9 runs (Fig.~\ref{fig:simres}), to investigate the effect of resolution on the disruption of satellites. We have focused on host haloes across three orders of magnitude in virial mass ($\MTWOC=[10^{12},10^{15}]$~$\msun$) and satellites of stellar mass $>10^{7}$~$\msun$. The simulations span a factor of $>8000$ in particle mass resolution, from $8.5\times10^4\msun$ (TNG50-1 aka TNG50) to $7\times10^8\msun$ (TNG300-3) in baryonic mass (Table~\ref{tab1}). We have contrasted populations of systems across different resolution runs but also, importantly, developed and used a set of halo matching catalogues to identify pairs of haloes and pairs of satellites between simulations of the same volume but at different resolution. 

Firstly, we studied properties of the host haloes, namely the subhalo/satellite mass functions and the stellar haloes / intracluster light profiles. We demonstrate that the subhalo mass functions are unchanged at the per~cent level by resolution (Fig.~\ref{fig:submf}), i.e. above the completeness limit where limited mass resolution clearly manifests itself ($>100$ DM particles per subhalo). On the other hand, the satellite stellar mass functions grow in amplitude at fixed stellar mass by {\it up to} a factor of two with each $2^{3}$ increase in the number of resolution resolution elements (Fig.~\ref{fig:satmf}). This is primarily due to increased star formation in satellites with improved resolution \mbox{\citep{Pillepich17b, Engler21}}, therefore mostly due to a shift along the $x$ axis of the mass function, rather than any difference in number of the satellites themselves \citep[see also a dedicated discussion for satellites of Milky Way-like galaxies in TNG50 by][]{Engler21b}. The stellar mass density radial profiles around galaxies likewise increase in amplitude by an average of 20~per~cent per step in resolution at all radii $<\RTWOC$ (Fig.~\ref{fig:sh}). The interpretation of quantitative comparisons with stellar halo observations will necessarily need to account for the impact of numerical resolution. Nevertheless, for the studied regimes, we discourage considering results from the low resolution runs of the TNG suite, i.e. we recommend that the reader focus on the flagship runs TNG50-1, TNG100-1 and TNG300-1 and on baryonic mass resolutions equal or better than $\sim 10^7\msun$. 

We next investigated the stripping of DM and stellar mass from satellites using pairs of satellites matched across simulations. We show that the stripping rate of DM is dominated by stochastic effects across all resolutions, host halo masses and satellite infall stellar masses (Figs.~\ref{fig:IEDM},~\ref{fig:STDM1},~\ref{fig:STDM2}). This is the case even down to satellites as-severely-stripped satellites as 10~per~cent of the initial, i.e. infall DM mass. In other words, the stripping rate of DM from satellites is realized consistently across all range of studied resolution levels as satellites are stripped of up to 90 per cent of their initial mass. The one exception is the lowest resolution simulation of the suite (TNG300-3) for which almost all satellites are disrupted at first pericentre. Some subhaloes also show an oscillating DM mass, which suggests that our results are affected by misattribution of the subhalo mass to the host halo around pericentre caused by the substructure finder. On the other hand, the resolution dependence of the stripping of stellar mass is much stronger, with higher resolution simulations typically retaining 50~per~cent of their infall stellar mass for 3~Gyr longer than their lower resolution pair counterparts (Fig.~\ref{fig:STiStar1}). However, most satellites still retain at least 10~per~cent of their stellar mass over the full 10~Gyr in which they are tracked, especially for the highest resolution pairing of TNG50-1 and TNG50-2 (Fig.~\ref{fig:STDM2}). We therefore show that the disruption times of TNG50-2 are not affected by resolution above the per~cent level, especially for $M_\rmn{*-Infall}>10^{8}$~$\msun$. Given that TNG50-2 has a similar resolution as TNG100-1, this means that the TNG100-1 flagship simulation can be considered converged at the per~cent level to TNG50-1 in the context of satellite stripping and disruption times.    

Finally, we analysed how matched pairs of satellites contribute stellar mass to their host haloes, focusing on galaxy groups and L* galaxy hosts. First, as per the effect already noted above, the stellar masses of satellites at infall are larger at better resolution, in both classes of host haloes: at TNG50-1 (TNG100-1) resolution, these are converged to better than the 30-50~per~cent (Fig.~\ref{fig:HSProgs}). We also see that poorer resolution causes satellites to contribute a larger proportion of their infall mass to the host, especially for massive satellites in group mass haloes. However, crucially, for simulations with mass resolution equal or better than about $\sim 10^7\msun$ (TNG300-1 or TNG100-2) there is no measurable difference across resolution levels, denoting a good level of convergence among the flagship runs (Fig.~\ref{fig:HSProgsRat}). Therefore, since the stellar concentration and stripping rate are converged but the total stellar mass is not, the subsequent ex-situ stellar mass increases with better resolution (Fig.~\ref{fig:HSDist}). This final figure also shows that the stellar mass stripped from satellites of higher resolution simulations is higher than in their lower resolution counterparts within the inner $0.1\RTWOC$ -- by more than 50~per~cent -- but outside this radius the high resolution contribution is still somewhat higher than for low resolution, at 40~per~cent to 20~per~cent. 

Based on these findings, we present the following overall picture for the formation of stellar haloes and ICL as a function of the simulation resolution, in the TNG model. First, better resolution leads to satellites at infall (and hosts) that have larger stellar masses. As the satellites are processed inside the host halo, poorer resolution simulations (TNG300-2, $m_\rmn{DM}>4.7\times10^{8}$~$\msun$) contribute a too large proportion of their mass to the outskirts of the host halo, but the satellites are not completely disrupted even after 10~Gyr; their elliptical orbits prevent the spurious disruption described for circular orbits by \mbox{\citet{vdBosch18}}. Better resolution simulations retain a higher fraction of the stellar mass deeper into the host halo, and so the time to strip to 50~per~cent of the infall mass is longer. However, because of their overall larger stellar masses, their satellites still deposit a larger absolute number of stars into the outer halo than do their lower resolution counterparts. The stripping times in the TNG50-2 simulation converge at the per~cent level compared to TNG50-1 and so, by extension, TNG100-1 is also robust to changes in resolution. Overall, the stripping rate decreases until the resolution level of TNG100-1, after which the stripping rate will not change significantly but the total amount of stellar mass will keep rising.

We have therefore shown that the TNG100-1 (aka TNG100) simulation is robust to resolution-dependent excess stripping of massive satellites at the per~cent level, but the rise in star formation efficiency nevertheless increases the mass of the extended stellar halo, an important caveat for quantitative comparisons with data. Finally, we note that we have not considered explicitly the contribution of very low mass satellites that cannot be resolved in TNG100-1 and TNG50-2 ($M_\rmn{*-Infall}<1.1\times10^{7}$~$\msun$); we expect that the total contribution from these galaxies is much smaller than their larger counterparts, but should be kept in mind nevertheless as part of any comprehensive study.    

\section*{Acknowledgements}

 We would like to thank the anonymous referee for a useful report. MRL acknowledges support by a Grant of Excellence from the Icelandic Research Fund (grant number 206930). AP acknowledges funding from the European Union (ERC, COSMIC-KEY, 101087822, PI: Pillepich). DN acknowledges funding from the Deutsche Forschungsgemeinschaft (DFG) through an Emmy Noether Research Group (grant number NE 2441/1-1). CE acknowledges support by the Deutsche Forschungsgemeinschaft (DFG, German Research Foundation) through project 394551440.

\section*{Data availability statement}
 All simulations of the IllustrisTNG suite i.e. all the TNG100, TNG300, and TNG50 simulation suites, are publicly available. Their data products can be downloaded from this URL (\url{https://www.tng-project.org/data/}) and directions for using the data can be found in \citet{Nelson19}. The halo matching catalogues published with this paper have been made available at \url{www.tng-project.org/lovell25}.  
 
 \bibliographystyle{mnras}

 \appendix
 
 \section{The dispersal of satellite orbits after infall}
 \label{app:orb}

The data in Figs.~\ref{fig:IEDM} and \ref{fig:IEStarmP} imply that the orbits of satellites change stochastically after pericentre. In this situation it is possible that one resolution simulation can scatter the satellite onto a large orbit that limits the stripping rate and the second resolution simulation instead scatters to a smaller orbit with a higher stripping rate: therefore, any resolution dependence of the halo disruption time may be washed out. However, in addition to the stochastic variation it is also plausible that the rate at which the orbits change is influenced by the difference in resolution, in which case larger changes in resolution will lead to larger scattering.  

\begin{figure}
     \centering
     \includegraphics[scale=0.33]{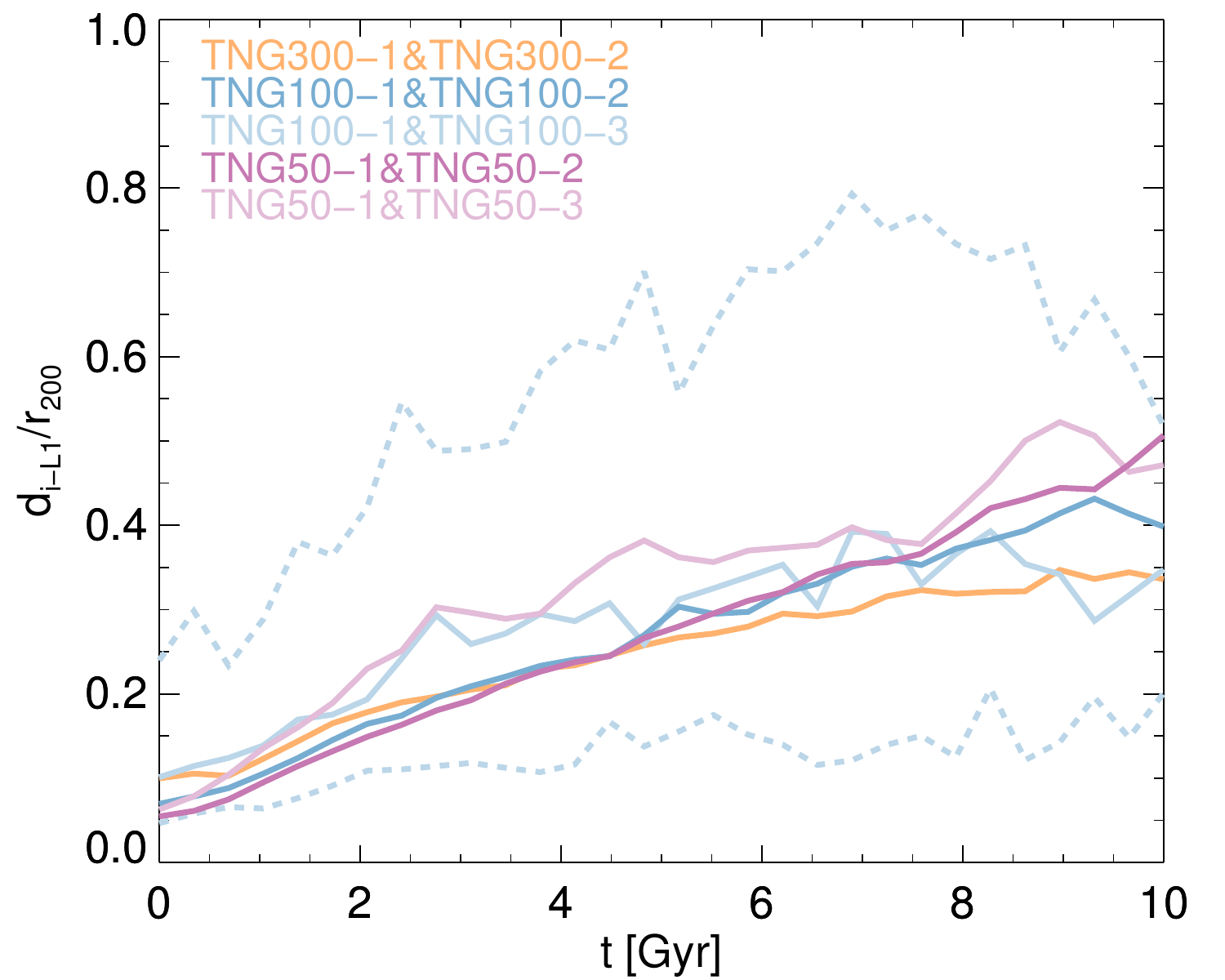}
     \caption{The median distance between satellite pairs after infall, measured in units of the host $\RTWOC$. TNG300-3 is omitted because most of its satellites are disrupted within 2~Gyr of infall. We include the 68~per~cent region for the TNG100-3 pair data as indicative of the rest of the data sets. The relationship between curve colour and data set is given in the figure legend. }
     \label{fig:DC}
 \end{figure}

We investigate the role of stochastic and systematic scattering by computing the distance between counterpart satellites as a function of time after infall. We include all satellites from all $M_\rmn{host}$-$M_\rmn{*-Infall}$ bins together, and compute the distances in units of the host $\RTWOC$ at that time, as opposed to $\RTWOC$ at the present day. We present the median relations as a function of time after infall in Fig.~\ref{fig:DC}; we omit TNG300-3 here because most of its satellites are destroyed after only 2~Gyr. In order to test whether the degree of scattering between resolutions is purely stochastic or instead has a systematic contribution, we plot the distance between L3 subhaloes and their L1 counterparts rather than L2: if the median partner distance between L1 and L3 counterparts is approximately the same as between L1 and L2 then the effect is purely stochastic, but if it is larger then there is also a systematic contribution to the orbit changes.

At infall, all of the data sets show the satellites at a median of $0.05\RTWOC$ apart, independent of volume or resolution. After even 0.5~Gyr divergence starts to occur, and the L3 to L1 partners show a faster divergence than L2 to L1, at a median distance of $0.3$~$\RTWOC$ after 4~Gyr, therefore the change in resolution does have a systematic impact on the rate of divergence. It takes approximately double that time for the L2 satellites to diverge by the same amount, but ultimately after 10~Gyr all of the pairs have diverged by a median of $0.4$~$\RTWOC$, and therefore it is no longer possible to compare the orbits of satellites on a case-by-case basis after some period of time. 

  \section{The contribution of post-infall star formation}
 \label{app:sb}

 In cases where a satellite is gas-rich at infall, the compression of this gas by its collision with the gas of the host halo can cause a starburst that dramatically increases the satellite stellar mass prior to and around the onset of stripping, in addition to less dramatic star formation events that occur in gas retained by the satellite after infall \citep{Vulcani18}. A comprehensive investigation of this effect is beyond the scope of this paper, and so we instead show the median post-infall star formation for our satellite sample to emphasise its importance in the context of stripping and then leave a more detailed analysis to future work.
 
\begin{figure}
     \centering
     \includegraphics[scale=0.335]{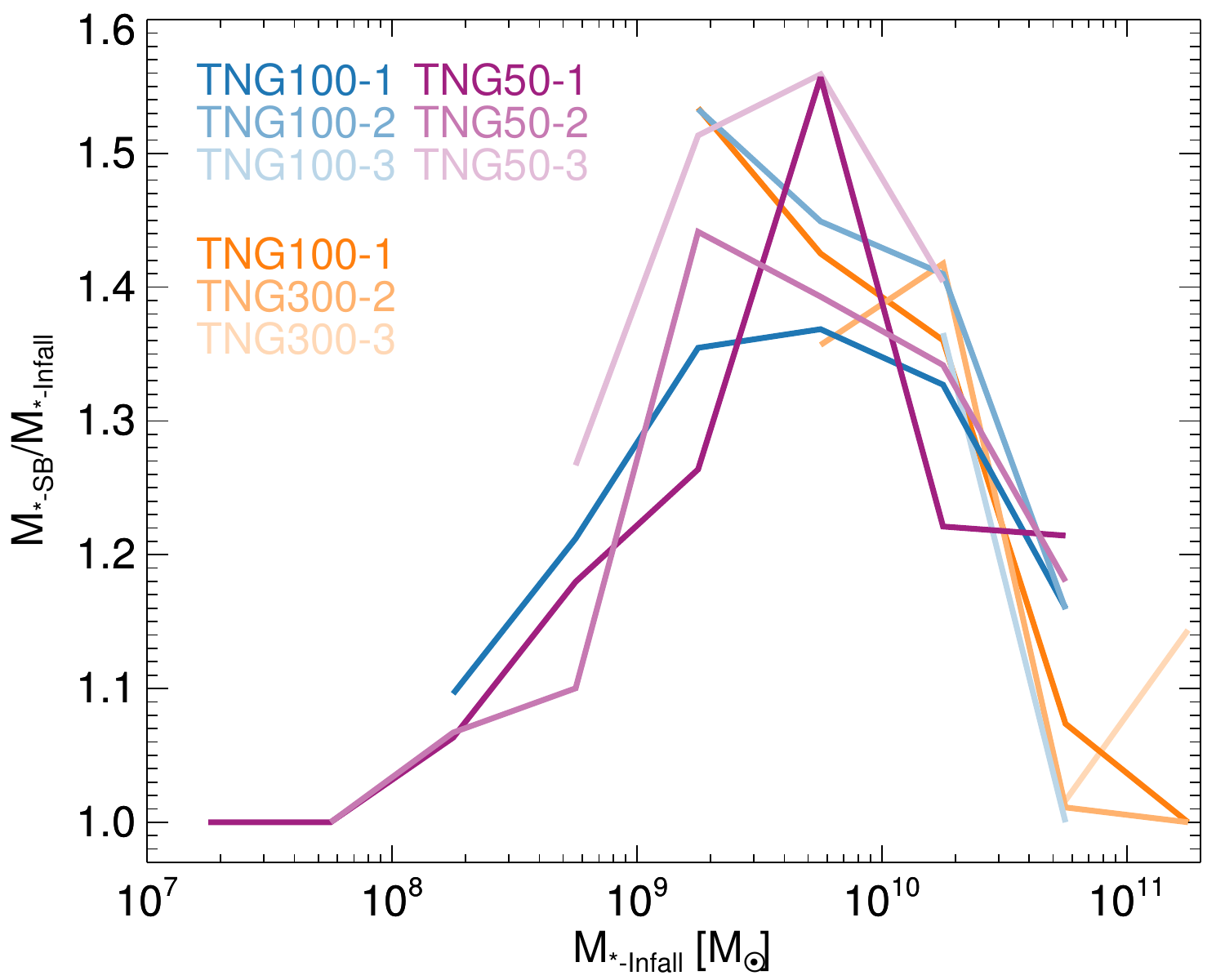}
     \caption{The median ratio of the maximum post-infall stellar mass to the infall stellar mass for our satellite sample, computed as a function of infall stellar mass. The correspondence between data set and line colour is given in the figure legend. }
     \label{fig:starburst}
 \end{figure}

 For each of our satellite data sets we compute the median ratio of the maximum stellar mass measured after infall, $M_\rmn{*-SB}$ to the infall stellar mass, $M_\rmn{*-Infall}$, and plot the result as a function of $M_\rmn{*-Infall}$ in Fig.~\ref{fig:starburst}. We include satellites that contain at least 100 star particles, and do not use separate bins for different host masses; note that a dedicated study would fully investigate the role of host mass.
 
 All of the models predict that satellites with $M_\rmn{*-Infall}\sim8\times10^{9}$~$\msun$ will on average increase their stellar mass by between 40~per~cent and 60~per~cent. In most of the simulations, the stellar mass gain falls away either side of $M_\rmn{*-Infall}\sim8\times10^{9}$~$\msun$, i.e. the size and incidence of post-infall star formation is lower. The decrease at higher masses likely reflects lower gas fractions driven by feedback from active galactic nuclei. At lower masses, supernova feedback may play a similar role, along with the shallower potential wells of lower mass satellites enabling the removal of gas by the ram pressure stripping. Finally, we cannot rule out that the ram-pressure stripping itself is resolution dependent: if, for example, the higher halo concentrations associated with higher resolution lead to greater resistance to ram-pressure stripping, the subsequent higher gas fractions may lead to still higher post-infall star formation.

\bsp	
\label{lastpage}
\end{document}